\documentclass[11pt, oneside]{article}   	
\usepackage{geometry}                		
\usepackage{amssymb,bm,amsmath,amsfonts}
\usepackage{multicol,lipsum,float}
\usepackage{graphicx}
\usepackage{subcaption}
\title{\textbf{Verification and validation of gyrokinetic and kinetic-MHD simulations for internal kink instability in DIII-D tokamak}}
\author{G. Brochard$^{1*}$, J. Bao$^{2*}$, C. Liu$^{3*}$, N. Gorelenkov$^3$, G. Choi$^1$, \\ G. Dong$^3$, P. Liu$^1$, J. Mc.Clenaghan$^4$, J. H. Nicolau $^1$, F. Wang$^{5}$, W. H. Wang$^1$, X. Wei$^1$, \\ W. L. Zhang$^2$, W. Heidbrink$^1$, J. P. Graves$^6$, Z. Lin$^1$, H. L\"utjens$^7$}
\date{}							
\begin{document}
\maketitle
\begin{center}
$^1$\emph{Department of Physics and Astronomy, University of California, Irvine, California 92697, USA} \\ $^2$ \emph{Institute of Physics, Chinese Academy of Sciences, Beijing 100190, China} \\ $^3$ \emph{Princeton Plasma Physics Laboratory, Princeton University, P. O. Box 451, Princeton, New Jersey 08543, USA} \\ $^4$\emph{General Atomics, PO Box 85608, San Diego, CA 92186-5608, USA} \\ $^5$ \emph{School of Physics, Dalian University of Technology, Dalian 116024, China}$^5$ \\ $6$ \emph{\'Ecole Polytechnique F\'ed\'erale de Lausanne (EPFL), Swiss Plasma Center (SPC), CH-1015 Lausanne, Switzerland } \\ $^7$ \emph{CPHT, CNRS, \'Ecole Polytechnique, Institut Polytechnique de Paris, Route de Saclay, 91128 Palaiseau, France}
\end{center}
\begin{center}
E-mail : gbrochar@uci.edu, zhihongl@uci.edu
\end{center}
* These authors contributed equally to this work
\begin{abstract}
\noindent
Verification and validation of the internal kink instability in tokamak have been performed for both gyrokinetic (GTC) and kinetic-MHD codes (GAM-solver, M3D-C1-K, NOVA, XTOR-K). Using realistic magnetic geometry and plasma profiles from the same equilibrium reconstruction of the DIII-D shot \#141216, these codes exhibit excellent agreement for the growth rate and mode structure of the internal kink mode when all kinetic effects are suppresed. The simulated radial mode structures agree quantitatively with the electron cyclotron emission measurement after adjusting, within the experimental uncertainty, the safety factor q=1 flux-surface location in the equilibrium reconstruction. Compressible magnetic perturbations strongly destabilize the kink, while poloidal variations of the equilibrium current density reduce the growth rate of the kink. Furthermore, kinetic effects of thermal ions are found to decrease the kink growth rate in kinetic-MHD simulations, but increase the kink growth rate in gyrokinetic simulations, due to the additional drive of the ion temperature gradient and parallel electric field. Kinetic thermal electrons are found to have negligible effects on the internal kink instability. 
\end{abstract}
\section{Introduction}
Large equilibrium currents in magnetically confined plasmas often excite magnetohydrodynamic (MHD) instabilities
including ideal kink modes \cite{Bussac1975} and resistive tearing modes \cite{Furth1963}, which can limit burning plasma performance and threaten fusion device integrity \cite{Hender2007}. These current-driven macroscopic instabilities have been extensively studied using MHD theory \cite{Bussac1975}\cite{Connor1985}\cite{Graves2000} and simulations \cite{Bondeson1992}\cite{Luetjens2001}\cite{Halpern2011}. However, the linear excitation and nonlinear evolution of MHD instabilities often depend on kinetic effects at microscopic scales and on the nonlinear coupling of multiple physical processes, e.g., microturbulence, neoclassical transport, and energetic particle effects. Therefore, fully self-consistent simulations of these current-driven MHD modes require a kinetic approach. Some kinetic effects such as the kinetic contribution to the energy principle's potential energy \cite{Kruskal1958}\cite{Chen1984} and the neoclassical polarization due to toroidally trapped particles \cite{Rosenbluth1998}\cite{Graves2000} and have been incorporated in kinetic-MHD simulations, with a fluid moment closure using current or pressure calculated from thermal and/or fast ion distribution function(s) \cite{Park1992}\cite{Todo1995}\cite{Briguglio1995}\cite{Kim2004}\cite{Brochard2020a}\cite{Liu2021}.\\\\Although linear and nonlinear kinetic effects are fully retained in the nonlinear gyrokinetic equation \cite{Brizard2007}, effects of equilibrium current have been neglected in most of gyrokinetic simulations \cite{Lee1983}. The gyrokinetic simulation model is suitable for describing low frequency plasma instabilities including microinstabilities \cite{Tang1978}, meso-scale Alfven eigenmodes excited by energetic particles \cite{Chen2016} and MHD modes driven by the equilibrium current and plasma pressure gradients \cite{Chen2016}\cite{Chen1991}. Recently, a gyrokinetic simulation model with equilibrium current \cite{Deng2012} has been implemented in the gyrokinetic toroidal code (GTC) \cite{Lin1998}, which was subsequently utilized for linear simulation of internal kink \cite{McClenaghan2014}, resistive \cite{Liu2014} and collisionless \cite{Liu2016} tearing modes, and drift-tearing modes \cite{Shi2019} in a cylinder or a high aspect-ratio tokamak with circular cross-section. Beside the kinetic contribution to the energy principle's potential energy and the neoclassical polarization in kinetic-MHD simulations, gyrokinetic simulations also contain effects of finite parallel electric field (e.g., mode conversion to kinetic Alfv\'en wave and driftwave instability drive due to thermal plasma pressure gradients) and off-diagonal terms of the pressure tensor. Furthermore, GTC has also incorporated compressible magnetic perturbations to maintain perpendicular force balance \cite{Dong2017}, which have been neglected in most of the previous gyrokinetic simulations. Recent GTC simulations found that these compressible magnetic perturbations can be important for interchange-like modes \cite{Dong2017}\cite{Choi2021} with realistic $\beta$ (ratio of kinetic pressure to magnetic pressure) in tokamak plasmas, consistent with analytic theory \cite{Berk1977}\cite{Tang1980}. \\\\ In this work, we initiate a verification and validation (V\&V) study \cite{Greenwald2010} for gyrokinetic and kinetic-MHD simulations of the current-driven internal kink instability in a real tokamak experiment. The verification focuses on a benchmark between a gyrokinetic turbulence code GTC \cite{Lin1998}\cite{Holod2009}, two kinetic-MHD initial value codes M3D-C1 \cite{Jardin2012}\cite{Ferraro2009}\cite{Liu2021} and XTOR-K \cite{Lutjens2008}\cite{Lutjens2010}\cite{Brochard2020a}, and two kinetic-MHD eigenvalue codes GAM-solver \cite{Bao2020}\cite{Zhao2021}\cite{Bao2021} and NOVA-K \cite{Gorelenkov1999}\cite{Gorelenkov2000}. The validation focuses on comparing the linear global simulation results of the current-driven instability with the experimental measurements in the DIII-D discharge \#141216. The dominant mode in this experiment has a toroidal mode number n=1 and appears first as a stationary internal kink mode, which later evolves into a fishbone mode with a frequency down chirping. All these five codes exhibit excellent agreement for the growth rate and mode structure of the n=1 internal kink mode when all kinetic effects are suppressed. The simulated radial mode structure agrees quantitatively with the electron cyclotron emission (ECE) measurements after adjusting, within the experimental uncertainty, the radial location of the safety factor q=1 flux-surface in the equilibrium reconstruction. Our study also finds that compressible magnetic perturbations strongly destabilize the kink, which is consistent with analytic theory \cite{Graves2019}, while poloidal variations of the equilibrium current density reduce the growth rate of the kink. Furthermore, kinetic effects of thermal ions decrease the kink growth rate in kinetic-MHD simulations, but increase the kink growth rate in the gyrokinetic simulations, due to the additional drive of the ion temperature gradient and parallel electric field. Finally, kinetic effects of thermal trapped electrons have negligible effects on the internal kink instability. \\\\We note that this is the first V\&V for the global gyrokinetic simulation of current-driven MHD modes in a real tokamak experiment. Previous V\&Vs for global gyrokinetic simulation focuse on meso-scales Alfv\'en eigenmodes excited by energetic particles \cite{Spong2012}\cite{Taimourzadeh2019} and microturbulence \cite{Xiao2015}. The validated global gyrokinetic simulations will enable integrated simulations incorporating multiple physical processes and treating both kinetic and fluid nonlinearities on an equal footing. Such integrated simulations of the nonlinear interactions between microturbulence, meso-scale Alfv\'en eigenmodes, and macroscopic kinetic-MHD modes are needed to predict realistically the confinement properties of energetic particles in burning plasmas \cite{Chen2016} \cite{Liu2021b} such as ITER. First-principles simulations effectively utilizing the most powerful supercomputers can also provide a large database for training reduced models and deep learning algorithms \cite{Dong2021}, which can be used for real time prediction and control of burning plasmas. \\\\ The rest of the paper is organized as follows. The target DIII-D experiment and its equilibrium reconstruction are described in Sec. 2. The verification for the internal kink simulations in the MHD limit is presented in Sec. 3. The validation of the MHD simulations against experimental ECE measurements is detailed in Sec. 4. Kinetic effects of thermal ions and electrons on the kink instability are discussed in Sec. 5. Finally, conclusions and discussions are given in Sec. 6.
\section{Description of the DIII-D discharge \#141216}
\begin{figure}[h!]
   \centering
   \includegraphics[scale=0.7]{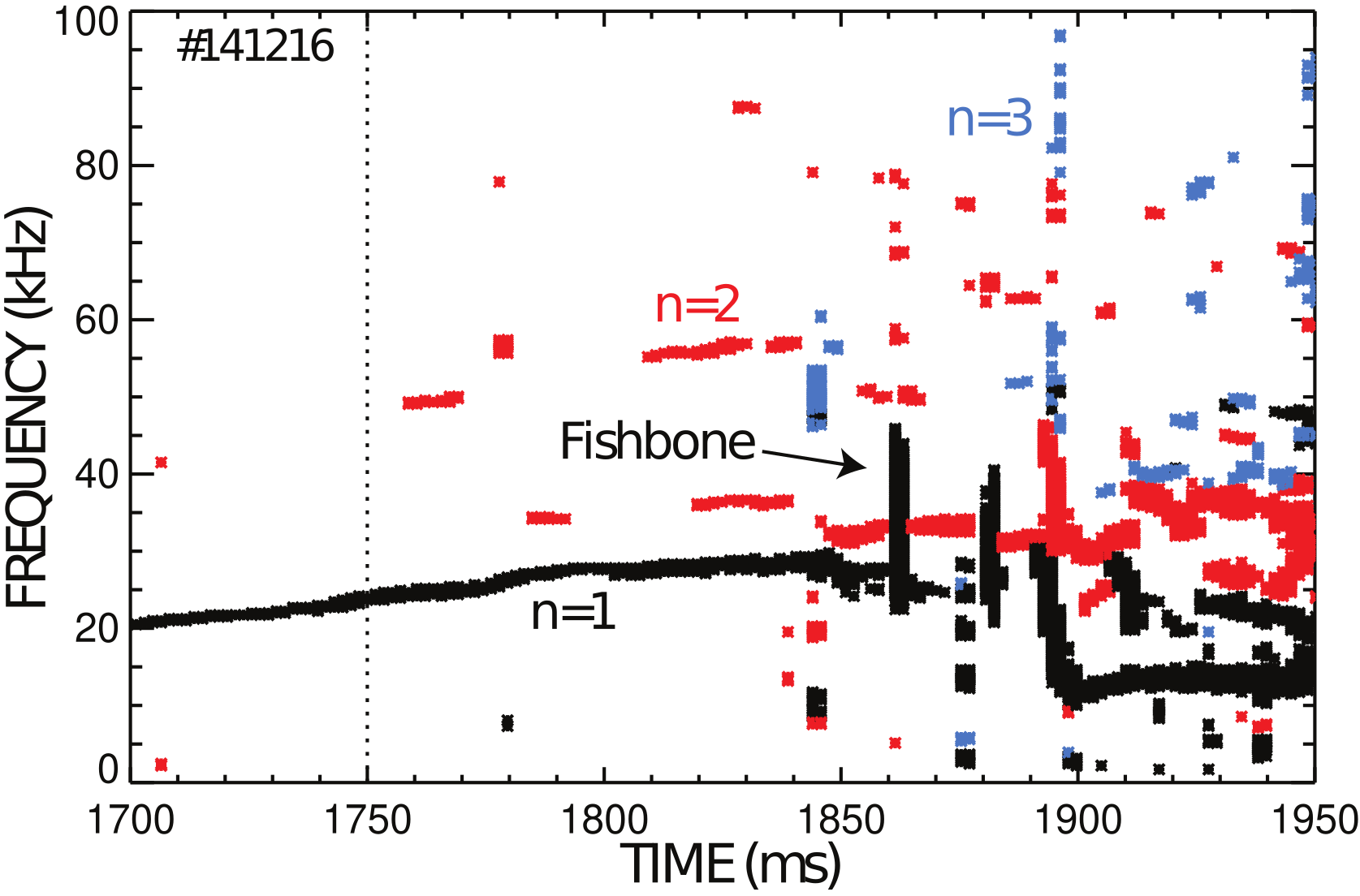}
   \caption{Experimental spectrogram in ms measured on the DIII-D discharge \#141216. The first toroidal harmonics from n=1 to 3 are respectively displayed in black, red and blue. A clear n=1 internal kink mode appears around t=1750ms, while a n=1 fishbone mode emerges at t$\sim$1890ms}
   \label{spec}
\end{figure}

The DIII-D discharge selected for analysis is a beam-heated, strongly shaped, H-mode plasma with ELMs.  After an ohmic phase, beam heating that begins at 1000 ms triggers an H-mode transition; then increasing beam power starting at 1640 s causes increasing MHD activities, including the kink activity at 1.75 s that is the focus of this study.  The spectrogram of the experimental discharge is displayed in Figure (\ref{spec}); the time evolution of a similar discharge appears in Fig. 6 of \cite{Heidbrink2015}. This spectrogram is obtained from the poloidal magnetic field fluctuations measured by multiple externally located magnetic probes, which resolve temporal-frequency and toroidal mode numbers.  The plasma shape is a lower single null divertor with elongation $\kappa=1.8$ and upper and lower triangularity of $\delta_u=0.44$ and $\delta_l=0.61$. The toroidal field is 2.0 T, the plasma current is 0.8 MA, and the safety factor at the surface that encloses 95\% of the toroidal flux is $q_{95}=6.8$.  The plasma is heated by 5.8 MW of deuterium neutral beams with injection energies of 75-81 keV that are all injected in the midplane in the direction of the plasma current. Plasma parameters at the magnetic axis include the electron density $n_e=5.10^{19}$ m$^{-3}$, electron and ion temperatures of $T_e=4.0$ keV and $T_i = 5.0$ keV, Zeff=1.5 with carbon being the dominant impurity. The toroidal rotation is 19 kHz at the q=1 position.  The ion temperature, toroidal rotation and carbon density are measured by charge exchange recombination spectroscopy of carbon impurities \cite{Gohil1990}. The toroidal rotation was not retained in the simulations, since the internal kink growth rate was found to be four higher than the experimental shearing rate at the q=1 surface.
\begin{figure}[h!]
\begin{subfigure}{.49\textwidth} 
   \centering
   \includegraphics[scale=0.22]{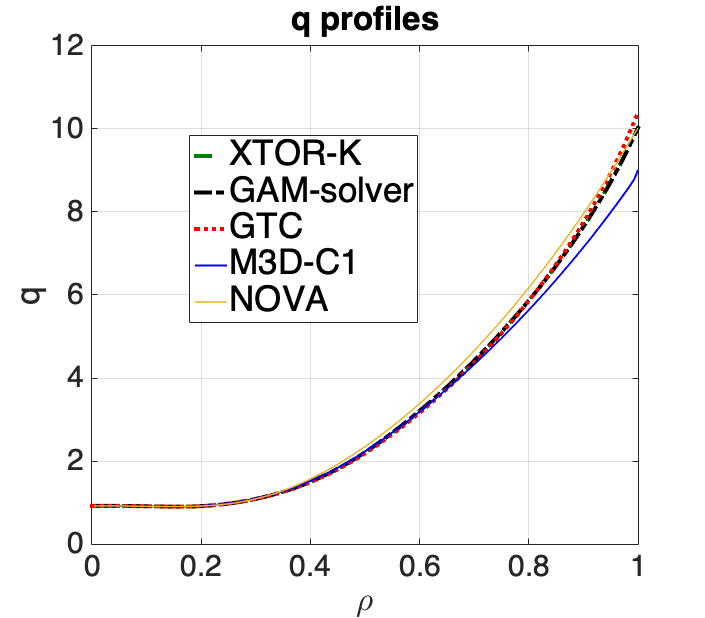}
   \caption{}
\end{subfigure}     
\begin{subfigure}{.49\textwidth} 
   \centering
   \includegraphics[scale=0.22]{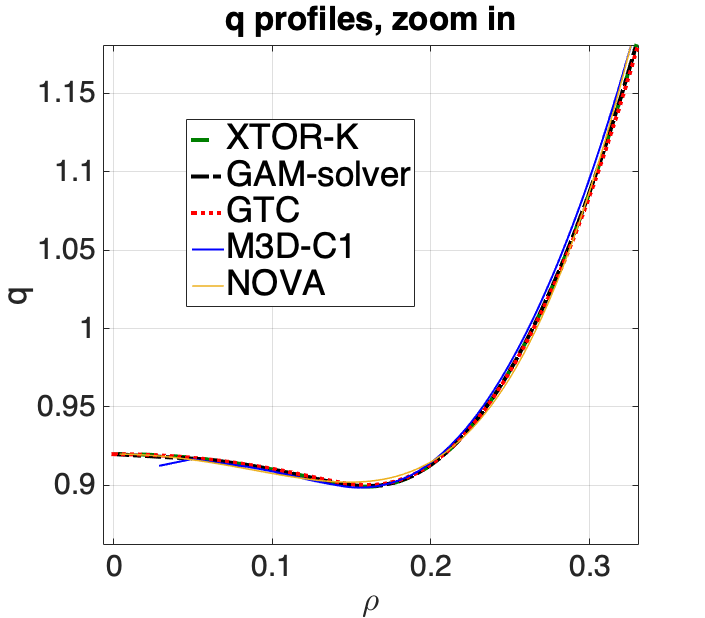}
   \caption{}
\end{subfigure}
\begin{subfigure}{.49\textwidth} 
   \centering
      \includegraphics[scale=0.18]{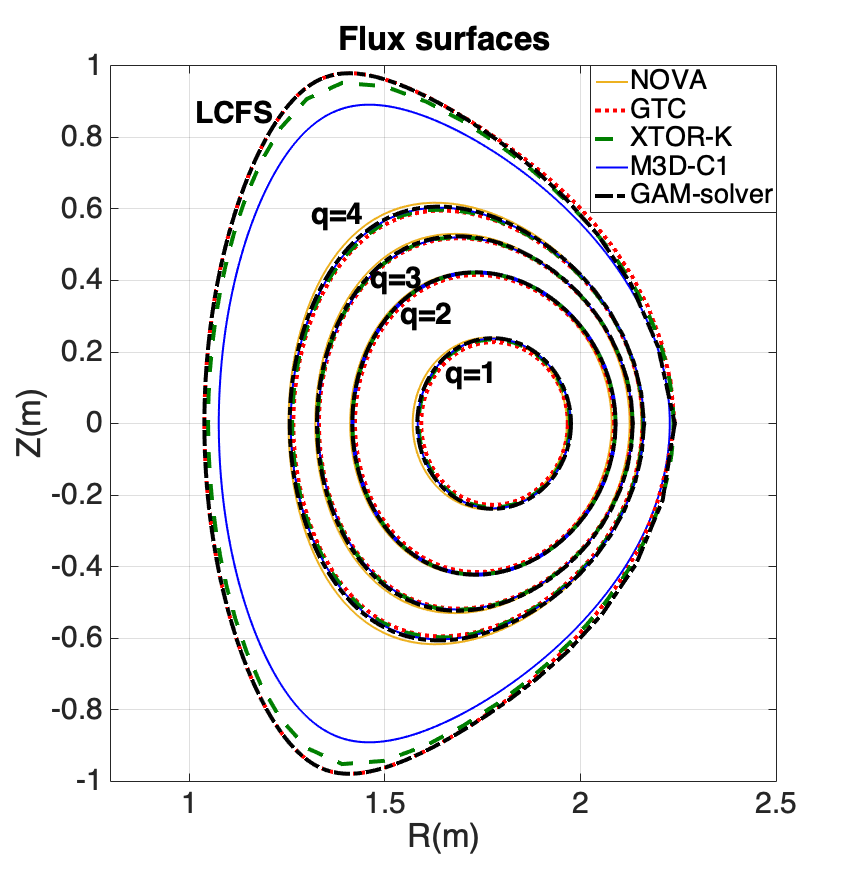}
   \caption{}
\end{subfigure}  
\begin{subfigure}{.49\textwidth} 
   \centering
      \includegraphics[scale=0.22]{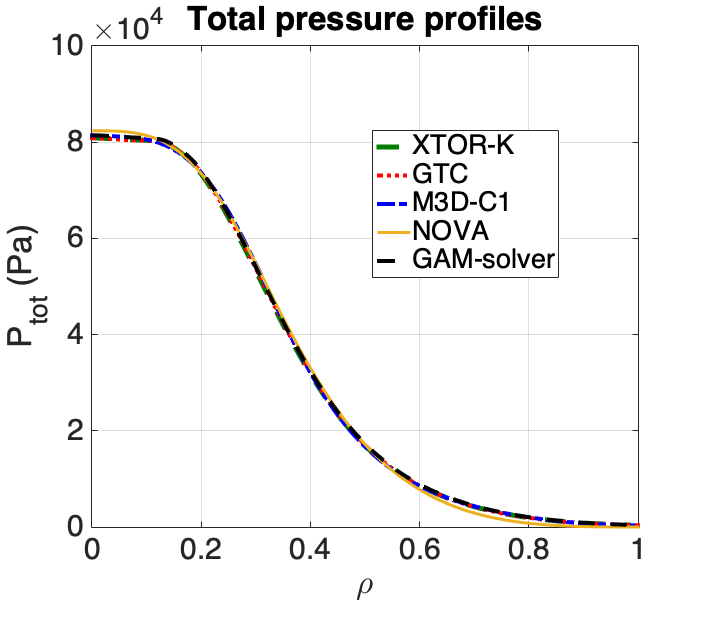}
   \caption{}
\end{subfigure}  
\caption{Features of the numerical equilibria used in all codes for the kink benchmark, reconstructed from the DIII-D shot \#141216 at t = 1750ms. (a) Safety factor profile (b) Safety factor profile in the core. (c) Flux surfaces in the poloidal plane. (d) Total plasma pressure in Pa.}
\label{MHD_eq}
\end{figure}
\\
The magnetic configuration of the DIII-D discharge \#141216 at $t=1750$ ms is reproduced with the kinetic EFIT code \cite{Lao1990}. The reconstruction includes internal magnetic field measurements obtained with MSE (Motional Stark Effect) to constraint the equilibrium. However, a post-treatment of the MHD equilibrium generated from EFIT is necessary for the code benchmark. First, some codes cannot use directly the EFIT output to run simulations, such as XTOR-K and NOVA-K. Secondly, small but finite differences in the q profile and the positions of the flux surfaces were noticed between the output of the different codes when using directly the EFIT equilibrium. Such differences would have led to quantitative mismatch when comparing the internal kink growth rates and mode structures obtained, given the sensitivity of the instability on such parameters. \\\\
Therefore, the EFIT MHD equilibrium has been used as input in the equilibrium code CHEASE \cite{Lutjens1996} to generate an uniform input for all codes. The time chosen for the discharge is in between two ELMs bursts, which is why there is no pressure pedestal for this H-mode plasma. We ensured that CHEASE and EFIT equilibria were similar by using the same safety factor profile, pressure profile and last closed flux surface. Equilibrium quantities as outputted from all codes are displayed in Figure (\ref{MHD_eq}). The radial coordinate used for the profiles is the square root of the normalized toroidal flux $\rho=\sqrt{\psi_T/\psi_{T,lcfs}}$, where $\psi_T$ is the toroidal flux and $\psi_{T,lcfs}$ the toroidal flux at the last closed flux surface. The pressure and safety factor profiles agree very well among all codes. In particular the position of the q=1 surface, the on-axis and minimum safety factor values $q_0$ and $q_{min}$ are exactly the same between all codes. The positions in the poloidal plane of the magnetic flux surfaces also match very well for the rational surfaces of q=1 to 4. Such an agreement allows us to perform a precise code verification for the internal kink instability in DIII-D.
\section{Verification of internal kink simulations in ideal MHD limit}
The code verification is performed using linear simulations from two perturbative eigenvalue codes (GAM-solver, NOVA-K) and three non-perturbative initial value codes, featuring one gyrokinetic code (GTC) and two kinetic-MHD codes (M3D-C1, XTOR-K). GTC is the only code implemented from a gyrokinetic formalism. The main features of each code are summarized in Table (\ref{tab1}). They are discussed in greater details with their respective simulation settings in appendix A.
\begin{table}[h!]
\center
\begin{tabular}{ |c| |c c c c c|}
\hline
 Code   & Formalism & Type & Fast ions & Thermal ions & Electrons \\ \hline \hline
   GAM-solver& Kinetic-MHD & Eigenvalue & Kinetic & Fluid & Fluid \\ 
   GTC & Gyrokinetic & Initial value & PIC GK, $\delta f$ & PIC GK, $\delta f$ & DKE fluid-kinetic  \\
   M3D-C1 & Kinetic-MHD & Initial value & PIC GK, $\delta f$ & Fluid/PIC GK, $\delta f$ & Fluid  \\
   NOVA-K & Kinetic-MHD & Eigenvalue & Kinetic & Fluid & Fluid  \\
   XTOR-K  &Kinetic-MHD & Initial value & PIC kinetic, full-f & Fluid/PIC kinetic, full-f & Fluid  \\ \hline 
 \end{tabular}
 \caption{Comparison of the simulation models used for the code verification. PIC stands for Particle in Cell, GK for gyrokinetic, DKE for drift kinetic equation, and $\delta$f/full-f stand for the simulation methods used to evolved respectively the perturbed or total distribution functions}
   \label{tab1}
\end{table}\\
The linear simulations are performed in the ideal MHD limit. For kinetic-MHD codes, this constitutes a natural limit of their full MHD model with plasma compressibility using an adiabaticity of 5/3. However for gyrokinetic codes, such a limit is more unusual. In GTC, it is achieved thanks to recent upgrades of the gyrokinetic simulation model incorporating the equilibrium current $J_{\parallel,0}$ \cite{Deng2012} and parallel magnetic compressibility $\delta B_{\parallel}$\cite{Dong2017}. When taking the limit of a massless electron fluid, neglecting all kinetic effects, and assuming no parallel electric field ($\phi_{eff}=0$, where $\textbf{E}_{\parallel} = -\nabla\phi_{eff}$), GTC gyrokinetic simulation model reduces to the incompressible ideal MHD model, which gives rise to the following dispersion relation \cite{Deng2012}\cite{Wei2021} 
\begin{equation}
\frac{\omega^2}{v^2_A}\nabla_{\perp}^2\phi + i\textbf{B}_0\cdot\nabla\bigg[\frac{\nabla_{\perp}^2(k_{\parallel}\phi)}{B_0}\bigg] + i\textbf{b}_0\times\nabla(k_{\parallel}\phi)\cdot\nabla\bigg(\frac{J_{\parallel,0}}{B_0}\bigg) - i\omega\mu_0\frac{\textbf{b}_0\times\boldsymbol{\kappa}}{B_0}\cdot\nabla\delta P = 0
\end{equation}
Here $\omega$ stands for the mode frequency, $v_A$ the Alfv\'en speed, $\phi$ the electrostatic potential, $B_0$ the equilibrium magnetic field, $\boldsymbol{\kappa}$ the magnetic curvature and $\delta P$ the perturbed plasma pressure. Note that this model incorporates the compressible magnetic perturbation $\delta B_{\parallel}$ to maintain perpendicular force balance, even though the perturbed flow is incompressible (i.e., no plasma compressibility). The effects of the compressible magnetic perturbation $\delta B_{\parallel}$, which has been neglected in most of gyrokinetic simulations, can be very important even for modest beta when simulating low frequency MHD modes such as interchange and kink instabilities for realistic $\beta$ in existing and future tokamak plasmas \cite{Berk1977}\cite{Tang1980}\cite{Graves2019}\cite{Choi2021}.\\\\
 To ensure that a quantitative agreement is met in the code benchmark, a $\beta$ scan of the internal kink growth rate is performed. The total pressure profile ranges from a fourth to the full EFIT pressure. For all cases considered the q profile is kept constant by modifying accordingly the current profile used as input in CHEASE equilibrium re-construction. This is done to vary as few significant parameters as possible, since the internal kink growth rate is very sensitive to the q profile shape. When only scanning the pressure profile, it is expected that the growth rate will monotonically increase with increasing pressure \cite{Bussac1975}, the mode being unstable past a threshold beta value.
\begin{figure}[h!]
   \centering
   \includegraphics[scale=0.37]{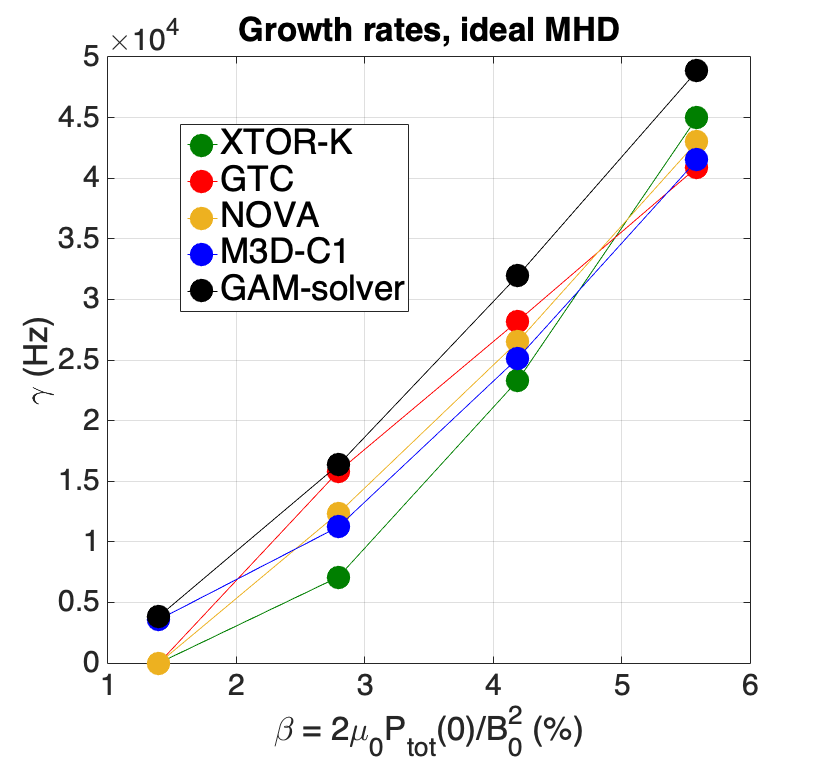}
   \caption{On-axis beta-scan of the internal kink growth rate obtained from the different codes, in the ideal MHD limit. An excellent quantitative agreement is obtained between codes using MHD and gyrokinetic formalisms. }
   \label{gamma_bench}
\end{figure}\\
The growth rates obtained from the beta scan are displayed in Figure (\ref{gamma_bench}). It can be observed qualitatively that for all codes, the growth rates do increase monotonically with the on-axis beta, and that the kink threshold is located near $\beta_{axis} \sim 1\%$. Quantitatively, an excellent agreement is obtained between all codes. The coefficients of variation $CV_{\gamma}= \sigma_{\gamma}/\mu_{\gamma}$ for each beta case are shown in Table (\ref{tab2}), where $\sigma_{\gamma}$ stands for the standard deviation and $\mu_{\gamma}$ the mean. The coefficient for the lowest pressure case is artificially high because this MHD equilibrium is close to the internal kink threshold, and therefore not included in Table (\ref{tab2}). Excluding this point, the coefficient average is 6.6\%, with 2.9\% for the experimental case. Since the ideal MHD limit is considered, the mode frequency is zero for all cases. In general, the internal kink mode frequency is the result of kinetic ion effects, such as their diamagnetic drift and the wave-particle resonance, which will be included in section 5. 
\begin{table}[h!]
\center
\begin{tabular}{ |c| |c c c|}
\hline
$\beta_{axis}$  & 2.8\% & 4.2\% & 5.6\%  \\ \hline 
   $CV_{\gamma}$  & 12\% & 4.9\% & 2.9\%  \\ \hline 
 \end{tabular}
 \caption{Coefficients of variations for the growth rates from all codes.}
   \label{tab2}
\end{table}
\begin{figure}[h!]
\begin{center}
\begin{subfigure}{.32\textwidth} 
   \centering
   \includegraphics[scale=0.23]{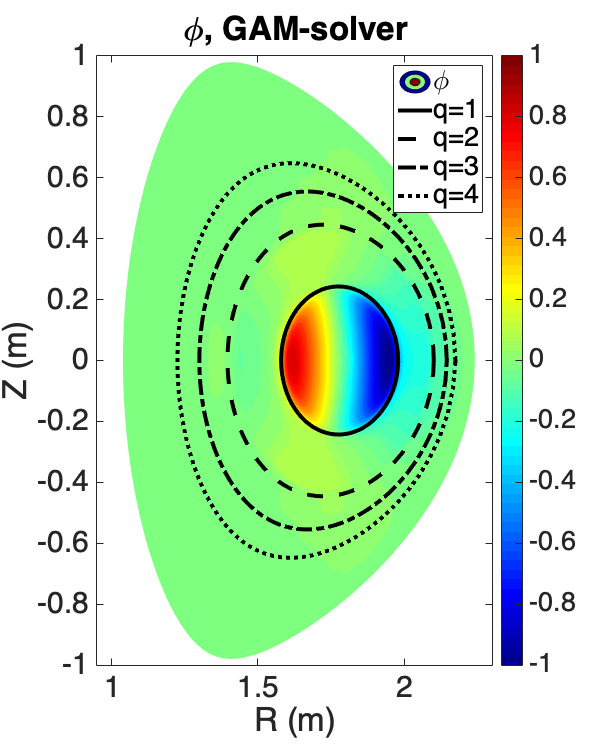}
   \caption{}
\end{subfigure}     
\begin{subfigure}{.32\textwidth} 
   \centering
   \includegraphics[scale=0.2]{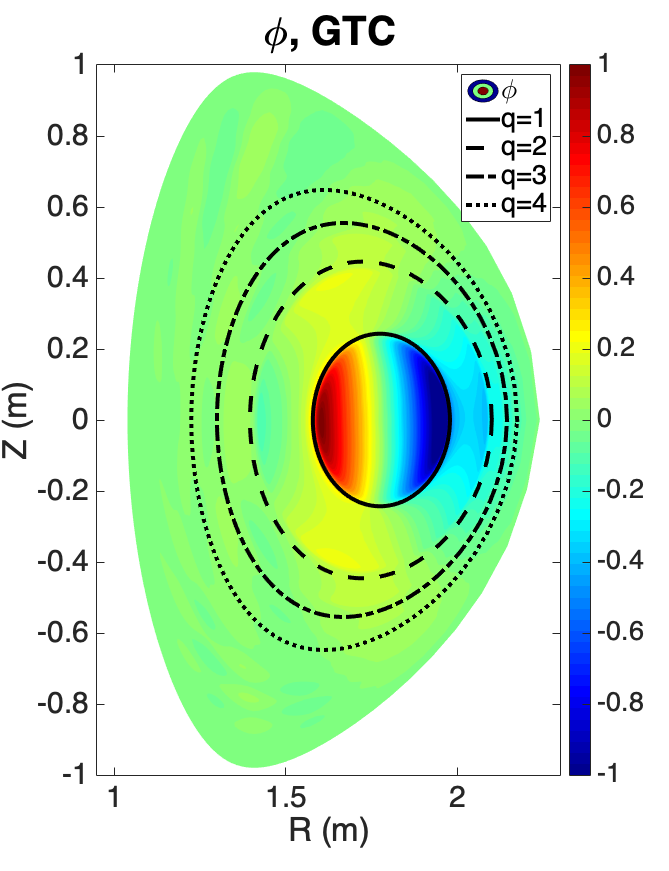}
   \caption{}
\end{subfigure}
\begin{subfigure}{.32\textwidth} 
   \centering
      \includegraphics[scale=0.2]{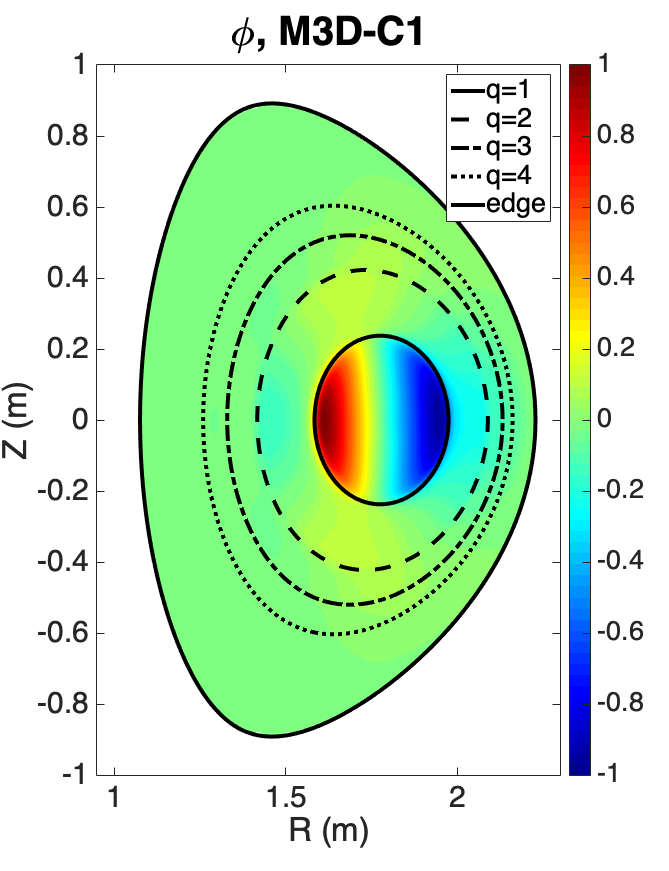}
   \caption{}
\end{subfigure}  
\begin{subfigure}{.32\textwidth} 
   \centering
      \includegraphics[scale=0.2]{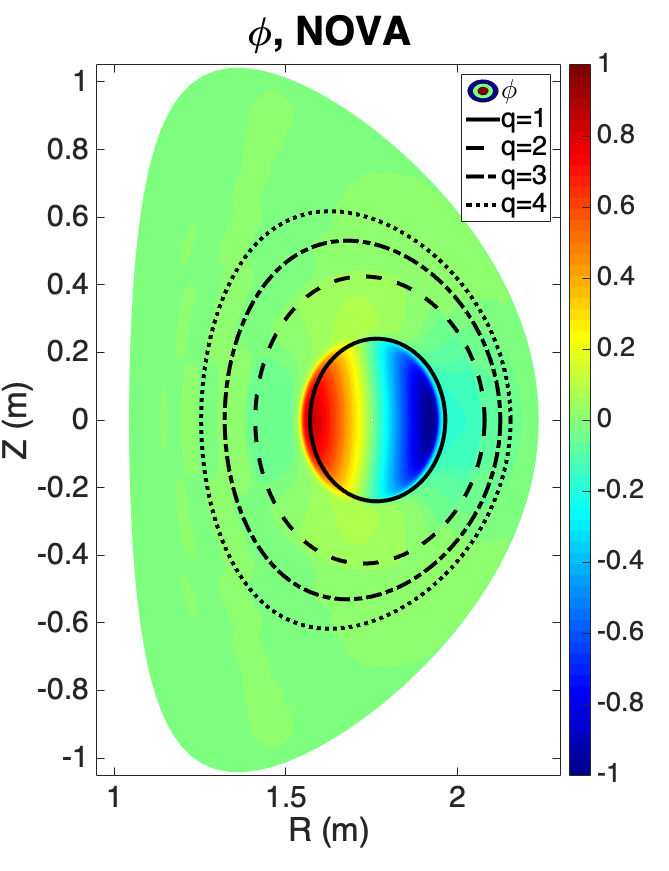}
   \caption{}
\end{subfigure}  
\begin{subfigure}{.32\textwidth} 
   \centering
      \includegraphics[scale=0.2]{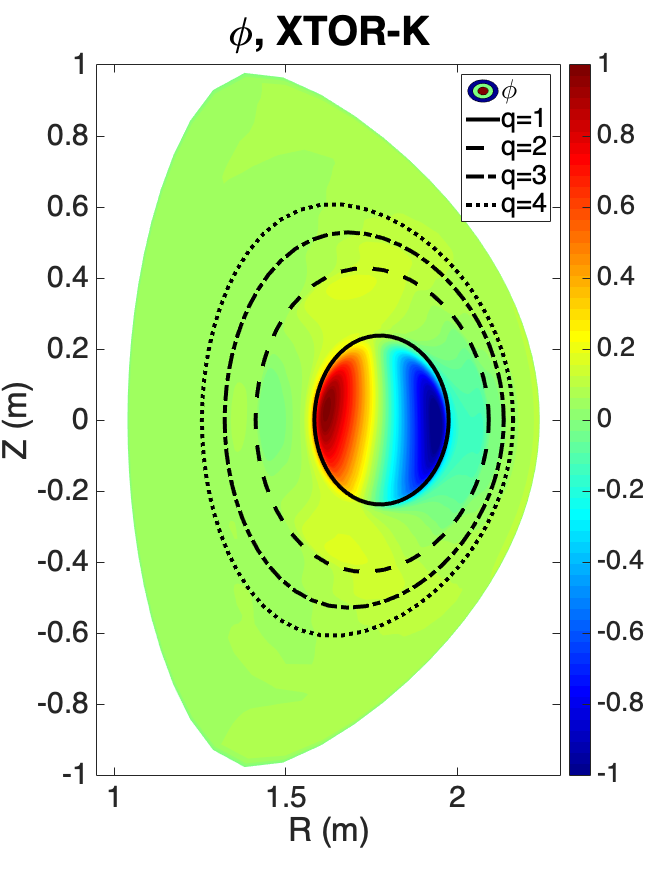}
   \caption{}
\end{subfigure}  
\caption{Mode structure in the poloidal plane for the electrostatic potential $\phi$ from the different codes. The black lines correspond to the flux surfaces from q = 1 to q = 4. The electrostatic potential mode structure agrees very well between all codes used in the benchmark.}
\label{mode2D_bench}
\end{center}
\end{figure}
\\To complete the code verification, the mode structure of the internal kink with the experimental pressure is compared between the codes in Figure (\ref{mode2D_bench}) and Figure (\ref{mode1D_bench}). Figure (\ref{mode2D_bench}) displays the electrostatic potential in the poloidal plane while Figure (\ref{mode1D_bench}) shows the poloidal harmonics of the electrostatic potential for the toroidal mode n=1. As it can be seen on these figures, an excellent agreement is obtained as well for the mode structures. In Figure (\ref{mode2D_bench}), a clear m=1 structure, well confined inside the q=1 surface, can be observed for all codes. The toroidal angle for each case has been adjusted to show the same phase between all codes. A subdominant m=2 harmonic, confined inside the q=2 surface, can also be identified in every simulation. These results are confirmed in Figure (\ref{mode1D_bench}), where it can be seen that the m=1 harmonic is indeed dominant. The m=1 electrostatic potential increases linearly until it peaks near q=1, which is consistent with linear theory. The position of the mode peak agrees well between the codes, with an average value of $\rho=0.23$ and a variation coefficient of 3.9\%. A subdominant m=2 harmonic does exist in all simulations, with a peak amplitude ranging between 10 to 20\% of the m=1 mode peak amplitude. Its radial profile varies significantly between the codes, potentially due to the different numerical grids and poloidal resolutions used between them (see appendix A).
\begin{figure}[h!]
\begin{center}
\begin{subfigure}{.32\textwidth} 
   \centering
   \includegraphics[scale=0.2]{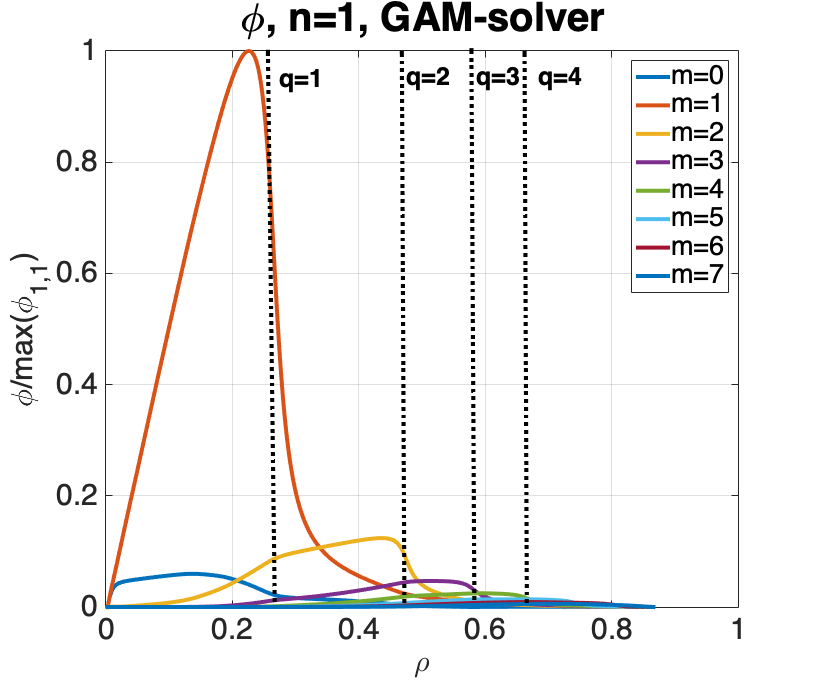}
   \caption{}
\end{subfigure}     
\begin{subfigure}{.32\textwidth} 
   \centering
   \includegraphics[scale=0.2]{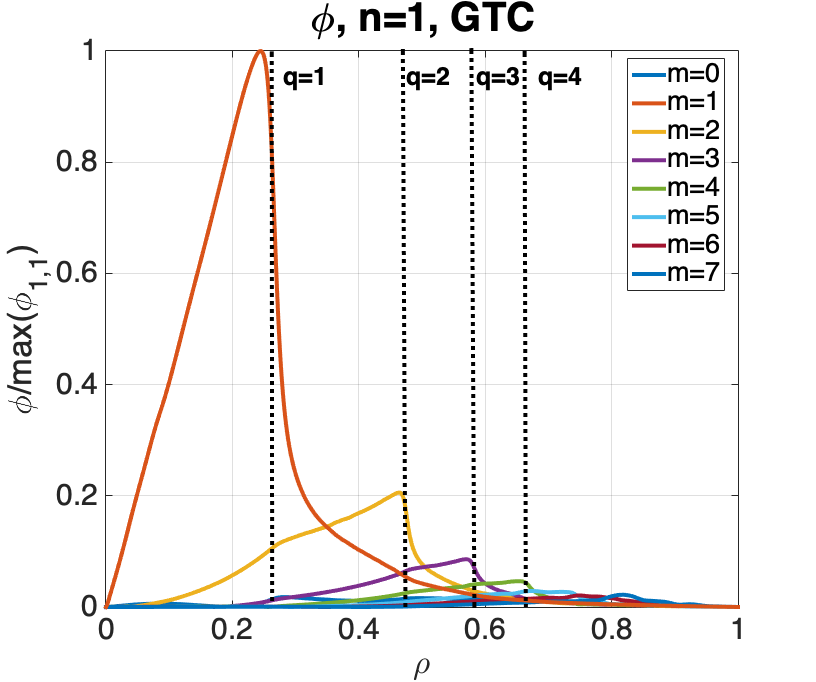}
   \caption{}
\end{subfigure}
\begin{subfigure}{.32\textwidth} 
   \centering
      \includegraphics[scale=0.2]{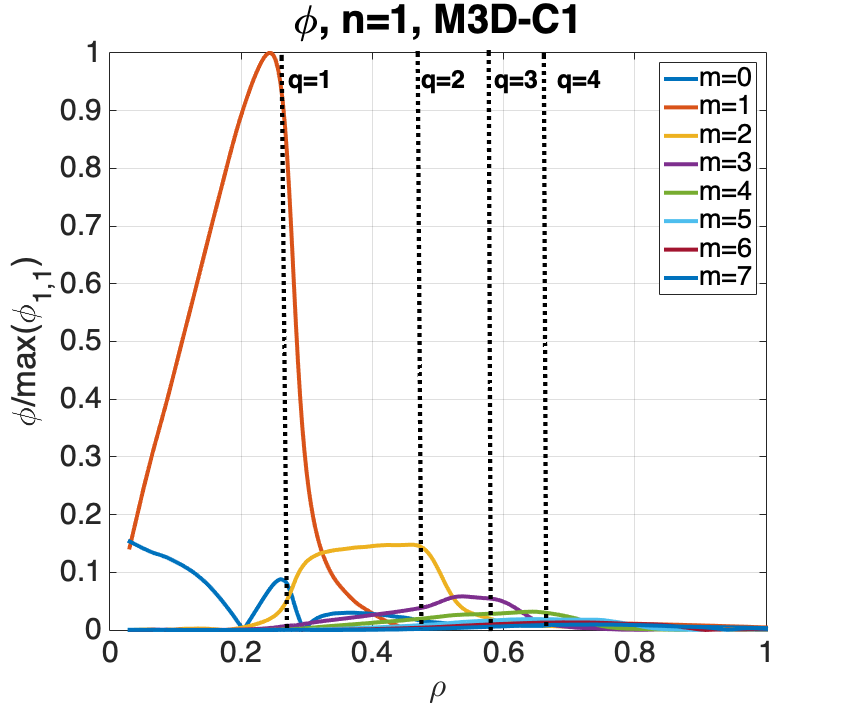}
   \caption{}
\end{subfigure}  
\begin{subfigure}{.32\textwidth} 
   \centering
      \includegraphics[scale=0.2]{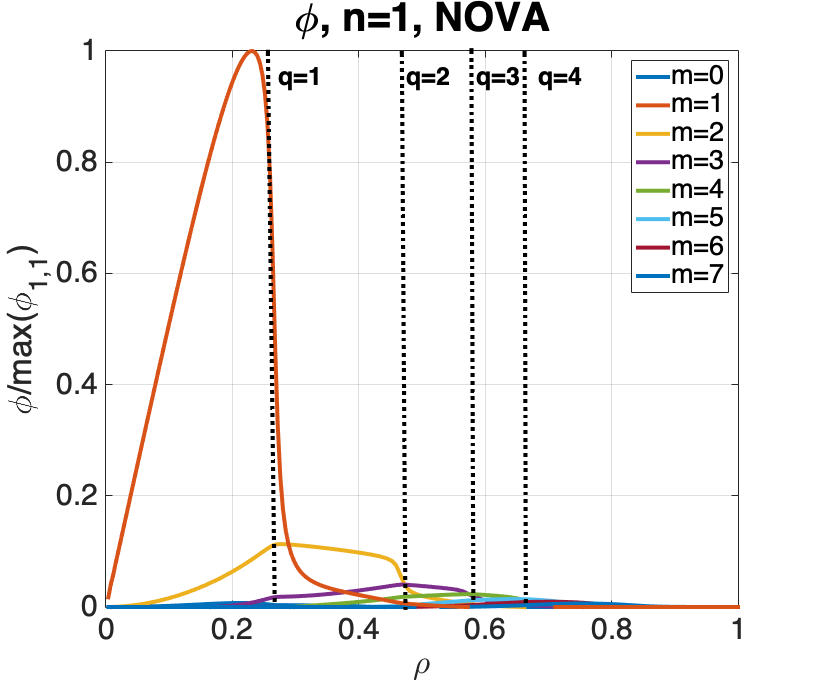}
   \caption{}
\end{subfigure}  
\begin{subfigure}{.32\textwidth} 
   \centering
      \includegraphics[scale=0.2]{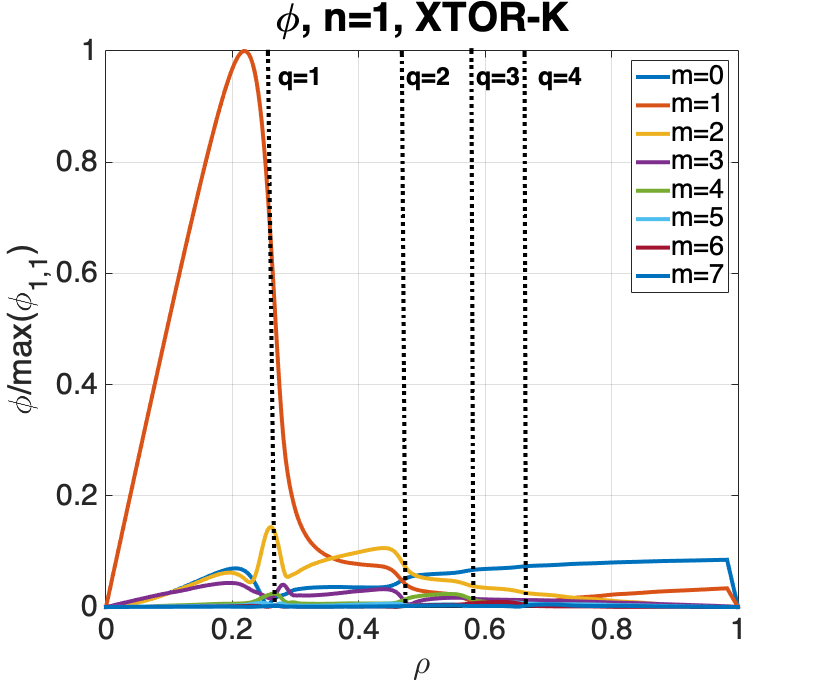}
   \caption{}
\end{subfigure}  
\caption{Radial mode structure of the electrostatic potential $\phi$ obtained from the different codes for different m harmonics, as a function of the normalized toroidal flux $\rho$. All harmonics are normalized by the maximum value of the m=1 harmonic. The structure of the m = 1 harmonic agrees very well between all codes used in the benchmark. The dashed lines correspond to the flux surfaces from q = 1 to q = 4. }
\label{mode1D_bench}
\end{center}
\end{figure}
\\Two key physical elements have been identified as essential to obtain a valid code verification. The first element is the accurate calculation of the equilibrium current that drives the kink instability. In Kinetic-MHD codes, the equilibrium current is directly obtained from the curl of the equilibrium magnetic field, without any approximations. Kinetic-MHD codes cannot run with inaccurate equilibrium current since the fields are not evolved perturbatively, the equilibrium fields need to respect the Grad-Shafranov equation explicitly. In gyrokinetic codes, the equilibrium current can also be calculated from the curl of the equilibrium magnetic field, which is typically taken as input from MHD equilibrium codes such as EFIT and CHEASE. However, the effects of the equilibrium current have been neglected (i.e., no kink drive) in most of the gyrokinetic codes, which typically only calculate perturbed current. For instabilities not mainly driven by equilibrium current, such an approximation is possible since only the perturbed fields are evolved in gyrokinetic simulations. Recently, GTC gyrokinetic simulations have incorporated effects of the equilibrium current in the simulations of the Alfv\'en eigenmodes \cite{Deng2012}, kink \cite{McClenaghan2014} and tearing \cite{Liu2014} modes. In GTC, the equilibrium current is directly derived from the curl of the magnetic field provided by EFIT or CHEASE, expressed in Boozer coordinates as
\begin{equation}\label{eq_cur}
\nabla\times\textbf{B}_0 = \frac{\partial g}{\partial\psi}\nabla\psi\times\nabla\zeta + \bigg(\frac{\partial I}{\partial \psi} - \frac{\partial \delta}{\partial\theta}\bigg)\nabla\psi\times\nabla\theta
\end{equation}
\begin{figure}[h!]
\begin{center}
\begin{subfigure}{.32\textwidth} 
   \centering
   \includegraphics[scale=0.2]{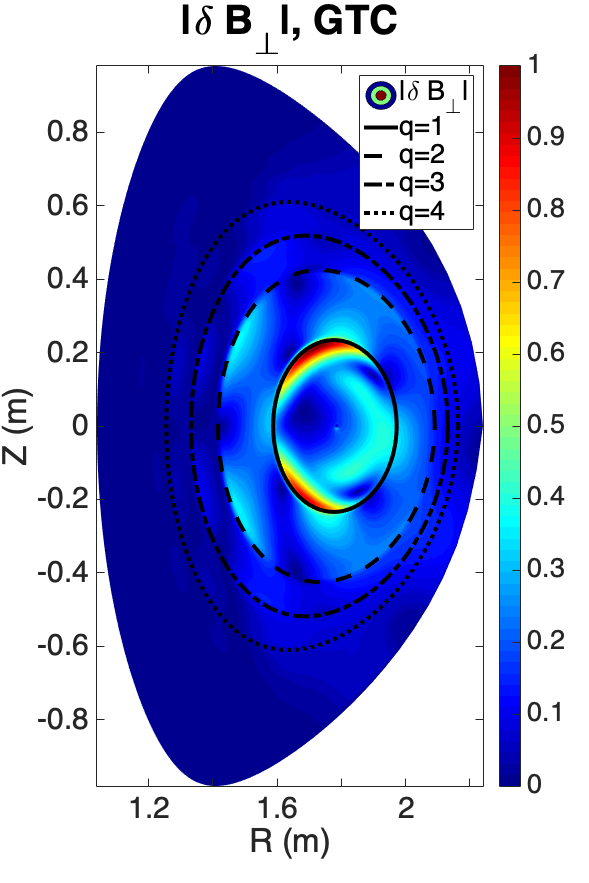}
   \caption{}
\end{subfigure}     
\begin{subfigure}{.32\textwidth} 
   \centering
   \includegraphics[scale=0.2]{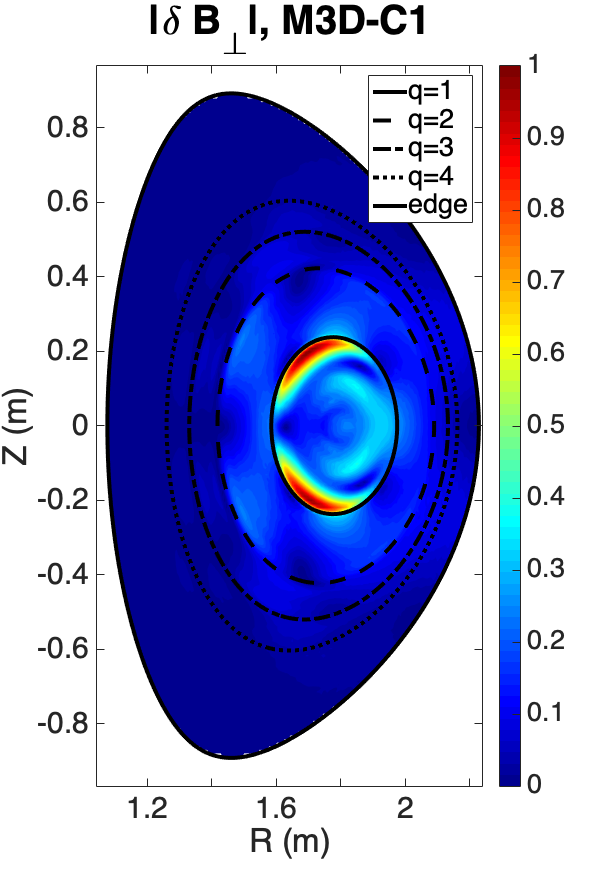}
   \caption{}
\end{subfigure}
\begin{subfigure}{.32\textwidth} 
   \centering
      \includegraphics[scale=0.2]{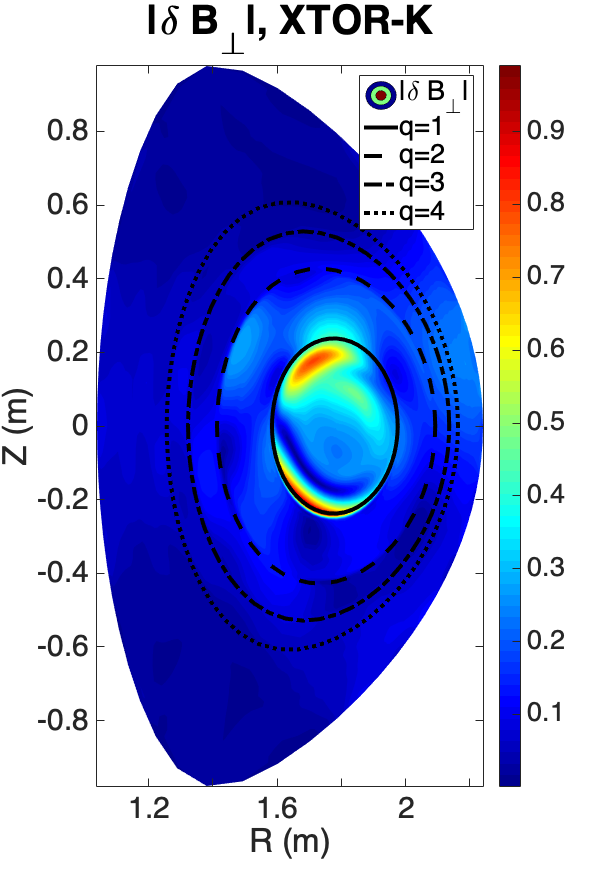}
   \caption{}
\end{subfigure}  
\begin{subfigure}{.32\textwidth} 
   \centering
      \includegraphics[scale=0.2]{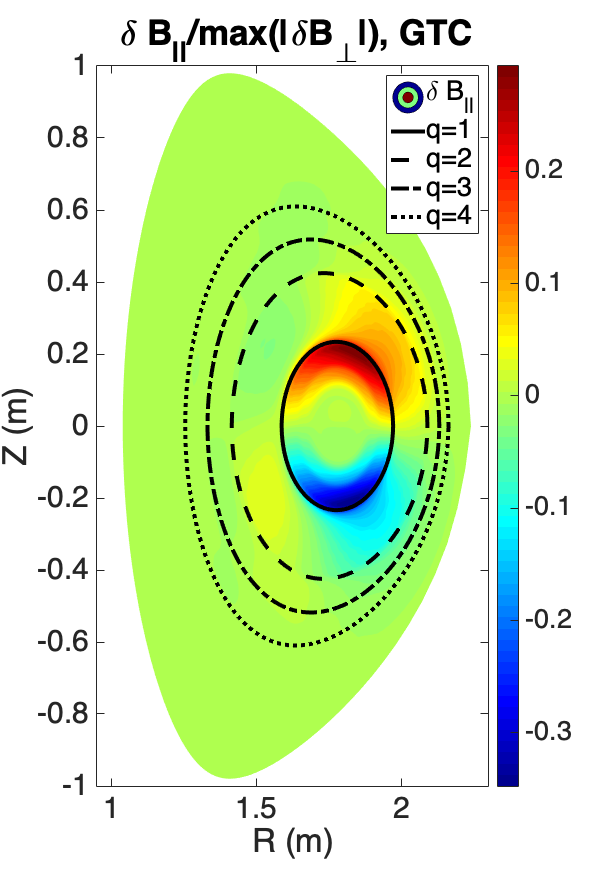}
   \caption{}
\end{subfigure}  
\begin{subfigure}{.32\textwidth} 
   \centering
      \includegraphics[scale=0.2]{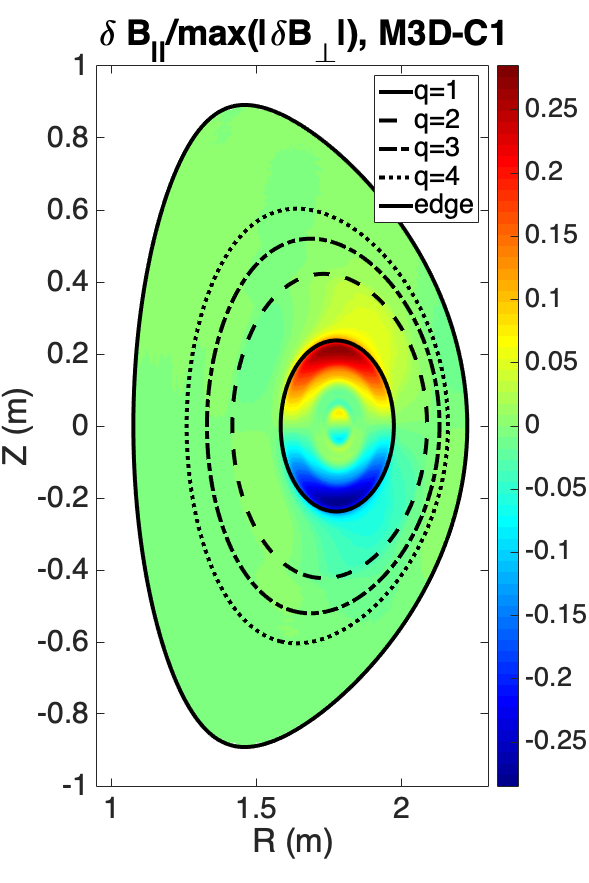}
   \caption{}
\end{subfigure}  
\begin{subfigure}{.32\textwidth} 
   \centering
      \includegraphics[scale=0.2]{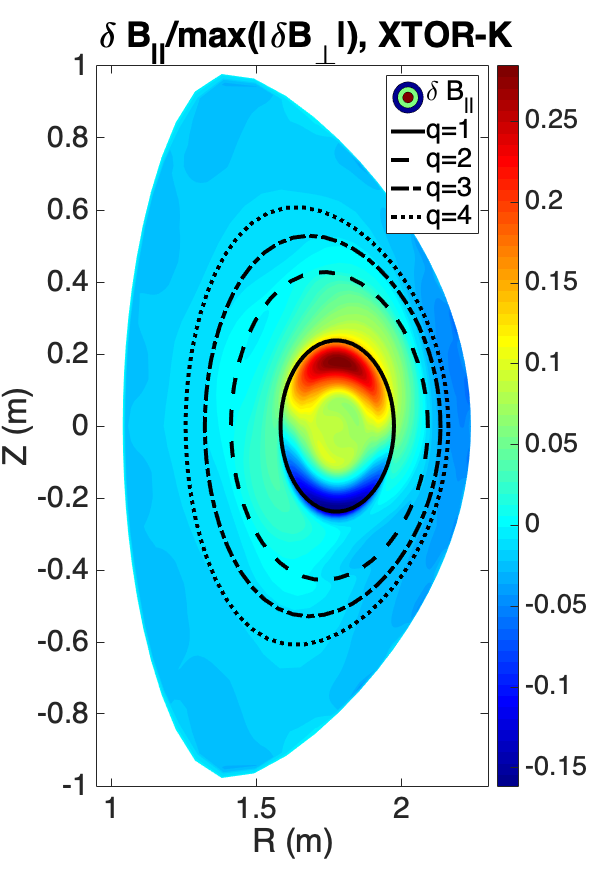}
   \caption{}
\end{subfigure}  
\caption{Mode structure in the poloidal plane for shear magnetic perturbation $|\delta B_{\perp}|$ (upper panels) and compressible magnetic perturbation $\delta B_{\parallel}$ (lower panels) from the different codes. The black contours correspond to the flux surfaces from q = 1 to q = 4. }
\label{dBpara}
\end{center}
\end{figure}where $g,I,\delta$ stand respectively for the covariant toroidal, poloidal and radial components of the magnetic field.  While $\delta=0$ in the cylindrical geometry and in the $s-\alpha$ model for the high aspect-ratio tokamak with concentric circular cross-section \cite{Deng2012}\cite{McClenaghan2014}\cite{Liu2014}, the component of the equilibrium current associated with the $\delta$ term in Eq.(\ref{eq_cur}) cannot be neglected in general geometry. It corresponds to an essential toroidal correction associated with the Shafranov shift, crucial for valid toroidal equilibrium. The term associated with $\delta$ in Eq.(\ref{eq_cur}) represents the poloidal variation of the equilibrium current, which maintains the self-consistency between the pressure and current drives for the MHD modes. We find that when this poloidal variation is artificially suppressed in GTC simulations, the internal kink growth rate increases by a factor of 4. The inclusion of the $\delta$ term in the equilibrium current is therefore essential for the code verification.\\\\
The second essential physical element is the inclusion of the compressible magnetic fluctuations $\delta B_{\parallel}$, to recover the limit of full MHD rather than reduced MHD which neglects $\delta B_{\parallel}$. Both GTC and M3D-C1 are able to operate with or without $\delta B_{\parallel}$, but these fluctuations are intrinsically built in the other codes. When $\delta B_{\parallel}$ is neglected, the internal kink mode is stable in both GTC and M3D-C1 simulations, for all on-axis beta cases. For this DIII-D experimental equilibrium, the on-axis beta is of order 5\%, which means that the $\delta B_{\parallel}$ effects can be very important even for this realistic $\beta$ value that is often considered as a low beta limit in many theory and simulations. In recent analytical work \cite{Graves2019}, the $\delta B_{\parallel}$ contribution is shown to be strongly destabilizing for the internal kink instability in tokamak plasmas with finite beta, to the extent that internal kink modes are unlikely to be excited in reduced MHD.  The numerical results obtained from this particular beta equilibrium in general geometry qualitatively agree with the analytical theory in the limit of circular flux surfaces and large aspect ratio. Therefore, compressible magnetic fluctuations cannot be neglected for the internal kink instability in general geometry for realistic beta plasmas. \\To further underline the importance of $\delta B_{\parallel}$ fluctuations in our simulations, the perturbed parallel and perpendicular magnetic fields are displayed in the poloidal plane in Figure (\ref{dBpara}). The mode structures of both perturbed quantities are found to be similar between all codes, and more importantly, the $\delta B_{\parallel}/$max($\delta B_{\perp}$) ratio is found to be also the same, about 25\%. 
\section{Validation of internal kink simulations in MHD limit}
Now that the code verification has been achieved, the simulation results obtained from XTOR-K and GTC are compared with the ECE measurements obtained from the DIII-D experiment. Electron temperature fluctuations are measured by a 40 channel radiometer that detects the second harmonic emission between 40-140 GHz \cite{Austin2003}.
\begin{figure}[h!]
   \centering
   \includegraphics[scale=0.27]{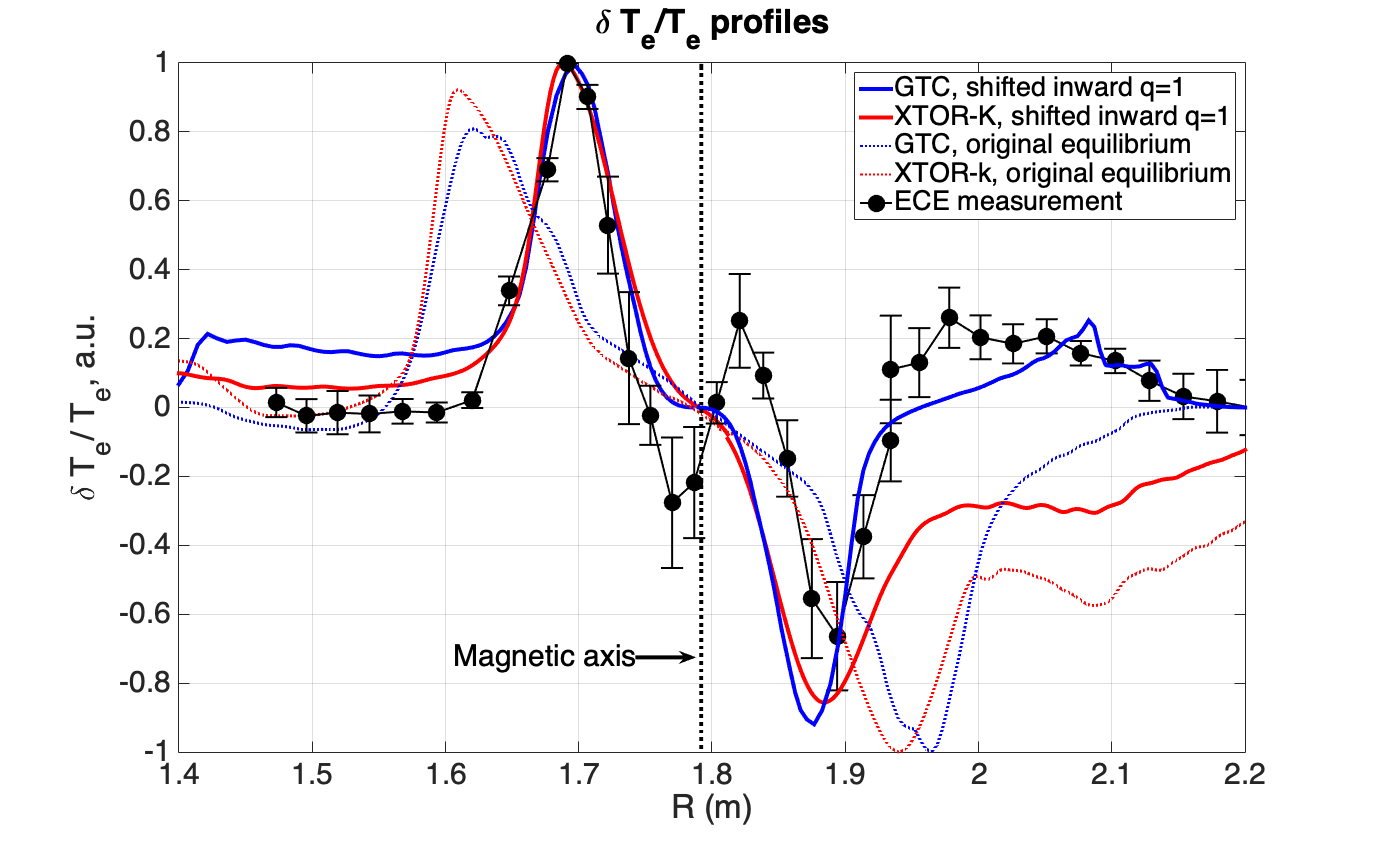}
   \caption{Normalized perturbed electron temperature $\delta T_e/T_e$ against the major radius R, obtained from GTC and XTOR-K MHD simulations (blue and red lines), and from the ECE measurement of the 21 kHz n=1 mode (black solid circle with error bar). Dotted lines correspond to MHD simulations using the original EFIT equilibrium, and solid lines to the modified EFIT equilibrium with the q=1 flux-surface shifted inward, see Figure (\ref{valid_comp}). The error bars are inferred from the random error in an ensemble of five Fourier transforms centered at 1750 ms.}
   \label{valid}
\end{figure}\\\\
As illustrated in Figure (\ref{valid}), the simulated $\delta T_e/T_e$ mode structures obtained from the original EFIT equilibrium agree qualitatively with the ECE measurement. $\delta T_e/T_e$ relates closely to the MHD displacement that can be expressed as $\xi_r=\delta T_e/ \partial_rT_e$. One should note here that the eigenfunctions from linear simulations are compared with the the nonlinearly saturated experimental results. The profiles are normalised by their maximum value to compare quantitatively the mode structures, without considering the saturation physics. A clear m=1 harmonic is present in each mode structure, with positive and negative perturbed electron temperature located asymmetrically across the magnetic axis. Satellite higher m harmonics can also be observed since the mode structure does not vanish after peaking near the q=1 surface. However, a quantitative agreement is not obtained between simulations and the experiment, since the mode structure peaks at different locations.\\\\
Given that the overall simulated mode structure seems to be only shifted compared to the experimental one, the quantitative mismatch probably comes from the uncertainty of the q=1 position in the reconstructed EFIT equilibrium. The EFIT equilibrium is reconstructed using magnetics, MSE, and kinetic pressure data.  The original equilibrium is in good agreement with the magnetics data (reduced chi-squared approximately unity), excellent agreement with the MSE data (reduced chi-squared well below unity), and good agreement with the pressure data. The computed flux surfaces have the same electron temperature on both sides of the magnetic axis. To obtain better agreement with the simulated mode structure, a new equilibrium is computed by moving the q=1 flux-surface slightly (within experimental uncertainty) inward to  $\rho=0.16$. This inward-shifted equilibrium is in better agreement with the magnetics data and equally good agreement with the pressure and Te data but the fit to the MSE data is relatively poor (reduced chi-squared $\sim 2$), so this reconstruction strains the limits of plausibility. Nevertheless, a new MHD equilibrium is created from CHEASE with this new q profile, keeping the last closed flux surface unchanged. However, the total pressure had to be changed as well because the internal kink instability was not excited using the original EFIT total pressure. With the original pressure profile and the modified q profile, the pressure gradient is almost flat in the entire q=1 volume, as displayed in Figure (\ref{valid_comp}) (b), which is not favorable to trigger the internal kink in analytical theory \cite{Bussac1975}. The pressure gradient therefore needs to be increased to excite the kink. 
\begin{figure}[h!]
\begin{subfigure}{.49\textwidth} 
   \centering
   \includegraphics[scale=0.21]{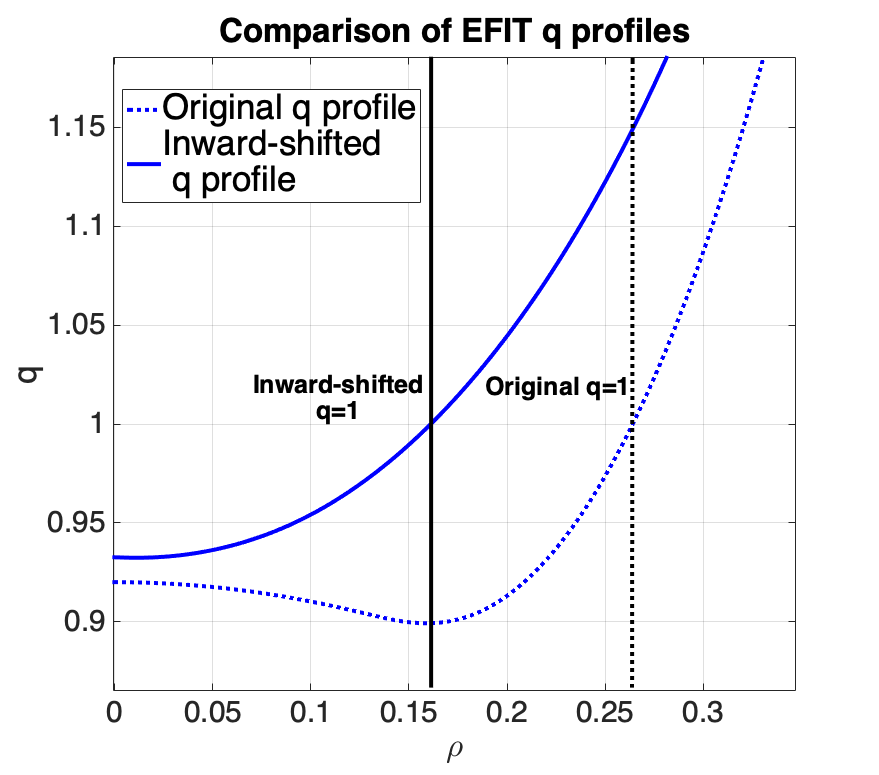}
   \caption{}
\end{subfigure}     
\begin{subfigure}{.49\textwidth} 
   \centering
   \includegraphics[scale=0.21]{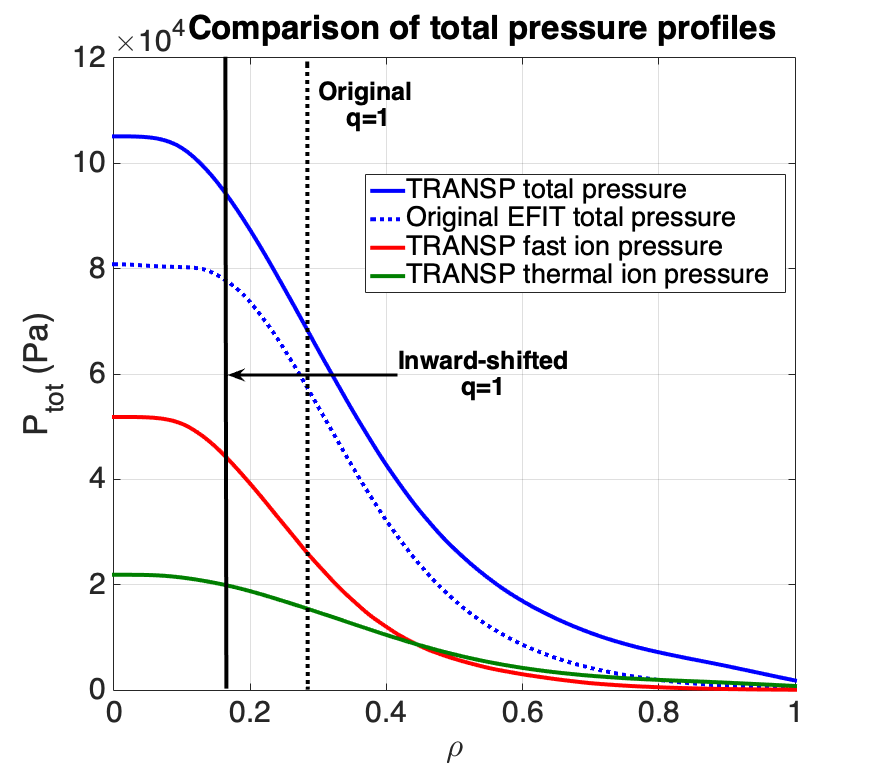}
   \caption{}
\end{subfigure} 
\caption{Comparison of (a) the q profiles and (b) the total pressure profiles between the original and inward-shifted EFIT reconstructions. The TRANSP total pressure is used with the inward-shifted q profile to excite the internal kink instability.}
\label{valid_comp}
\end{figure}\\\\
As shown in Figure (\ref{valid_comp}) (b), more than half of the total pressure comes from fast ions. Given that the experimental measurement errors on the fast ion pressure are non-negligible at the core plasma, the on-axis total pressure profile may have been underestimated in the original equilibrium reconstruction. Instead of the original EFIT pressure profile, the TRANSP total pressure is used for the new MHD equilibrium, which corresponds to the upper bound of the experimental pressure, which represents the sum of the pressures from the different plasma species.  The new MHD equilibrium is then another physical alternative to the original one, a certain range of equilibria being plausible due to the non-negligible error bars of the measurements at the core plasma.\\\\
Using the modified MHD equilibrium, a quantitative agreement is obtained between the simulated and the experimental mode structures. On the high field side the agreement is excellent, while there is a slight mismatch for the mode peak on the low field side, still acceptable considering the error bars on the ECE measurement. Towards the plasma edge the mode structure differs between the simulations and the experiment. Finite mode structure near the edge is due to small but finite higher m harmonics.\\\\
Both gyrokinetic and kinetic-MHD  codes have therefore been validated in the limit of ideal MHD against ECE measurement from DIII-D. As expected from linear theory, the internal kink is very sensitive to parameters such as the pressure and q profiles in the core. The experimental uncertainties in this zone make the internal kink difficult to reproduce from a traditional EFIT reconstruction. Kinetic-MHD and gyrokinetic simulations can therefore also be used to provide a more precise localization of some key parameters for the internal kink instability such as the position of the q=1 flux-surface.
\section{Kinetic effects of thermal plasmas on internal kink instability}
The verification and validation of kinetic-MHD and gyrokinetic codes in the ideal MHD limit constitutes a necessary step before kinetic effects can be studied in gyrokinetic simulations of macroscopic instabilities. So far, large scale kinetic-MHD instabilities such as the fishbone mode have mostly been analysed numerically thanks to nonlinear codes relying on a MHD formalism \cite{Fu2006}\cite{Vlad2016}\cite{Brochard2020b} . In these codes, some kinetic effects can be added by including pressure/current moments in the MHD momentum equation. A comparison with results obtained from a gyrokinetic formalism is therefore interesting since the kinetic-MHD formalism does not include the kinetic Alfv\'en wave and the drift-wave physics when two-fluid effects are not retained in the MHD equations.\\\\
In this paper, comparisons between initial values kinetic-MHD codes (M3D-C1, XTOR-K) and gyrokinetic codes (GTC) are presented. Since two-fluid effects and off-diagnonal terms of the pressure/current tensor are not considered in the kinetic-MHD codes, they do not incorporate the kinetic Alfv\'en wave and the drift-wave physics and do not retrieve some Landau damping effects of kinetic particles due to the absence of a finite parallel electric field in Ohm's law. These effects are present by default in GTC simulations, where the ion gyrokinetic equation \cite{Brizard2007} is solved using the particle method \cite{Lee1983}. The electron drift kinetic equation is solved using the fluid-kinetic hybrid model \cite{Holod2009}\cite{Lin2001}, where the perturbed electron distribution function is expanded order by order. In the lowest order, electron response is adiabatic and thus reduced to the massless fluid. Electron kinetic effects are then incorporated in the higher order non-adiabatic responses. \\ The physical models used between the gyrokinetic code and kinetic-MHD codes with kinetic effects are therefore different, which implies that results may qualitatively differ. Moreover, out of simplicity, isotropic Maxwellian distribution functions were used to describe kinetic species. While this is reasonable for thermal ions and electrons, this is not the case for energetic particles. The fast ion distribution in this DIII-D discharge was obtained via co-passing beam injection; it then departs significantly from an isotropic one. As a result, kinetic effects of fast ions have not been included in this study. They will instead be included in our next V\&V work on the fishbone phase of this DIII-D discharge, with realistic slowing-down distributions.
\begin{figure}[h!]
\begin{subfigure}{.49\textwidth} 
   \centering
   \includegraphics[scale=0.25]{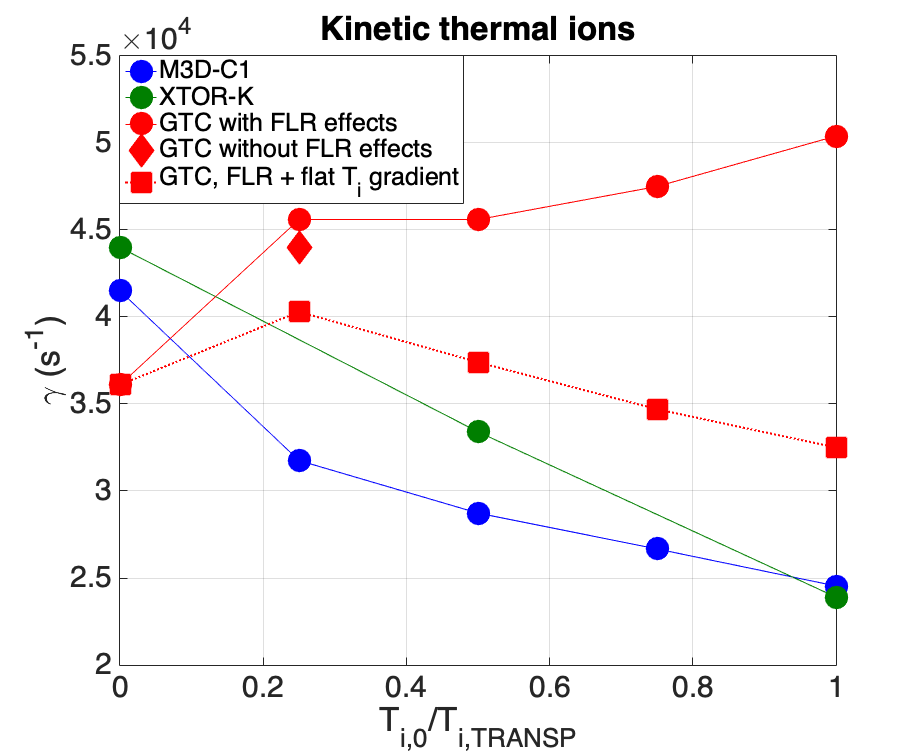}
   \caption{}
\end{subfigure}     
\begin{subfigure}{.49\textwidth} 
   \centering
   \includegraphics[scale=0.25]{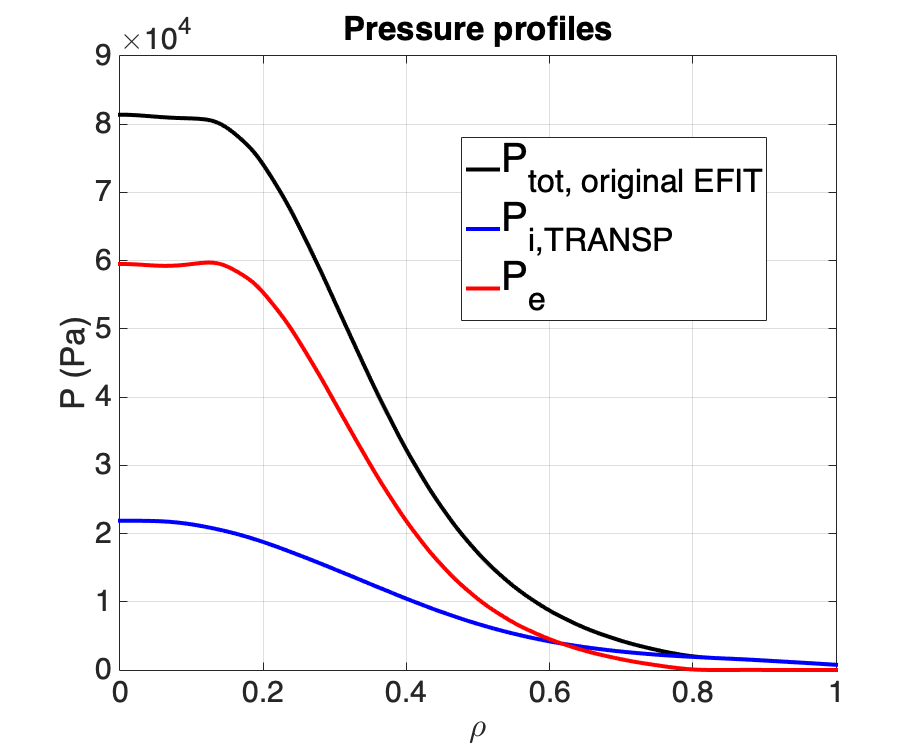}
   \caption{}
\end{subfigure} 
\caption{(a) Thermal ion temperature scan for the internal kink growth rate obtained from M3D-C1, XTOR-K and GTC. (b) Pressure profiles for thermal electrons and ions. The thermal ion pressure is taken from TRANSP simulations, the electron pressure profile is defined such that the sum of electron and ion pressures adds up to the total pressure obtained from original EFIT pressure.}
\label{gamma_GKi}
\end{figure}\\\\
Kinetic effects of bulk ions and electrons on the internal kink instability are included one after the other in order to isolate their individual effects. XTOR-K treats kinetic ions with a 6D phase-space dynamics, while M3D-C1 and GTC use a 5D guiding center model. Thermal ion density and temperature profiles are obtained from experimental measurements. A thermal ion temperature scan for the internal kink growth rate is performed, as displayed in Figure (\ref{gamma_GKi}) (a), ranging from $T_i=0$ up to the measured ion temperature. In each case, the electron temperature is adjusted so that the total plasma pressure is kept unchanged. This allows to single out thermal ion kinetic effects. Profiles for the ions and electrons are provided in Figure (\ref{gamma_GKi}) (b). Considering the on-axis parameters for thermal ions and electrons, $n_{i,0}=n_{e,0}=4.88\times 10^{19}$m$^{-3}$, $T_{i,0}=2.8$ keV, $T_{e,0}=7.6$ keV, $Z_{eff}=1.48$, $m_{i}=2m_p$ where $m_p$ is the proton mass, the ion-ion collision frequency is $\nu_{ii} =9.7\times10^2$ s$^{-1}$. For the same parameters, the average bounce frequency of thermal ions can be estimated as $\omega_{b,i} \sim 4.0\times 10^4$ s$^{-1}$. The thermal ions therefore lie in the banana regime since $\nu_{ii}\ll\omega_b$. Since electron temperature is higher than the ion temperature, the electrons also lie in the banana regime. No collision effects are taken into account in both gyrokinetic and kinetic-MHD simulations. Toroidal plasma rotation is not retained in all simulations, since the equilibrium shearing rate is four times lower than the internal kink growth rate in the ideal MHD limit.
\\\\
As illustrated in Figure (\ref{gamma_GKi}) (a), kinetic thermal ions have stabilizing effects on the internal kink instability in kinetic-MHD simulations. M3D-C1 and XTOR-K agree both qualitatively and quantitatively on the kink growth rate. This stabilizing effect is probably due to trapped thermal ions that are known to provide a stabilizing contribution to the energy principle through both the $\delta W_k$ term \cite{Chen1984} and the ion inertial enhancement \cite{Graves2000} (or neoclassical polarization \cite{Rosenbluth1998}). Diamagnetic effects from thermal ions can also contribute to the stabilization. On the other hand, gyrokinetic simulations find that kinetic thermal ions have a destabilizing effect with increasing temperature. A possible explanation could be that the thermal ion gradient brings an additional drive to the internal kink mode, as it does for the drift-wave instabilities. This effect is only present in the gyrokinetic formalism incorporating driftwave and kinetic shear Alfv\'en wave physics associated with finite parallel electric field. It can also be noted that FLR effects provide a finite destabilization effect with increasing ion temperature.\\\\To verify this additional instability drive by the ion temperature gradient, GTC simulations with flat thermal ion temperature are therefore performed, while keeping the total pressure profile unchanged. A temperature scan with uniform ion temperature is displayed in Figure (\ref{gamma_GKi}) (a). It can be seen that with a flat ion temperature profile, gyrokinetic thermal ions have a stabilizing effect on the internal kink, which is similar to the results obtained in kinetic-MHD simulations. A quantitative agreement cannot be achieved because with the flat ion temperature profile, the stabilizing contributions from the $\delta W_k$ term and the diamagnetic rotation are lowered since they are proportional to the thermal ion pressure gradient. The change in $T_i$ cannot be compensated by a change in $n_i$ since the ion density profile is required to be the same as the electron density profile to preserve quasi-neutrality. Yet, the qualitative agreement between gyrokinetic and kinetic-MHD simulations indicates that the same physical mechanisms lead to kink stabilization are present in both formalisms. Even with this flat temperature profile and in the limit of $T_i \rightarrow 0$, the kink growth rate from the gyrokinetic simulation is slightly larger than that from the ideal MHD simulation, which is most likely due to the finite parallel electric field (i.e., kinetic shear Alfv\'en wave). These results show that the inclusion of drift-wave and kinetic shear Alfv\'en wave physics is necessary to capture all the kinetic effects in the dynamics of macroscopic instabilities.  \\\\ In these gyrokinetic/kinetic-MHD simulations, the internal kink has a finite mode frequency for all cases, but it is much smaller than the growth rate, and therefore too low to be measured with these initial value codes in reasonable simulation time. 
\begin{figure}[h!]
\begin{subfigure}{.49\textwidth} 
   \centering
   \includegraphics[scale=0.32]{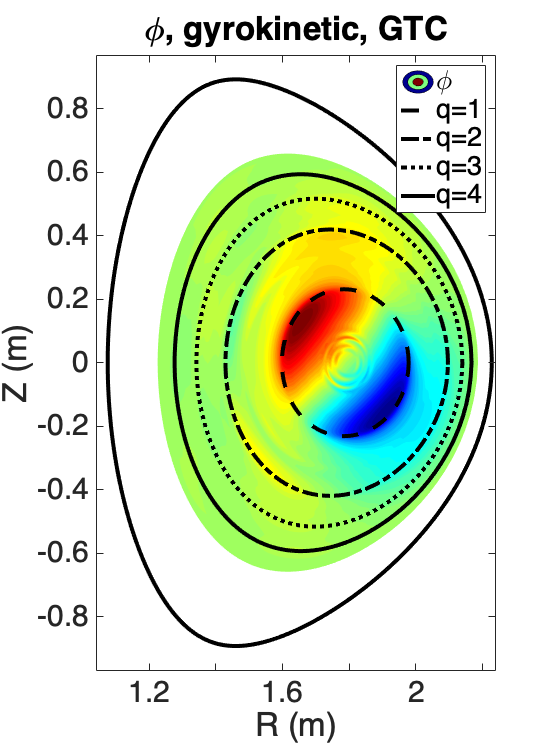}
   \caption{}
\end{subfigure}     
\begin{subfigure}{.49\textwidth} 
   \centering
   \includegraphics[scale=0.28]{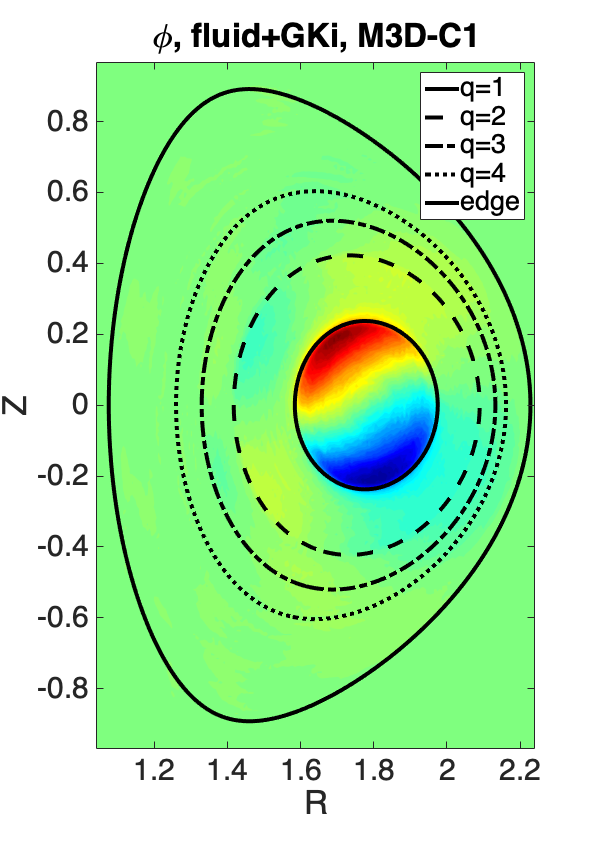}
   \caption{}
\end{subfigure}
\begin{subfigure}{.49\textwidth} 
   \centering
   \includegraphics[scale=0.27]{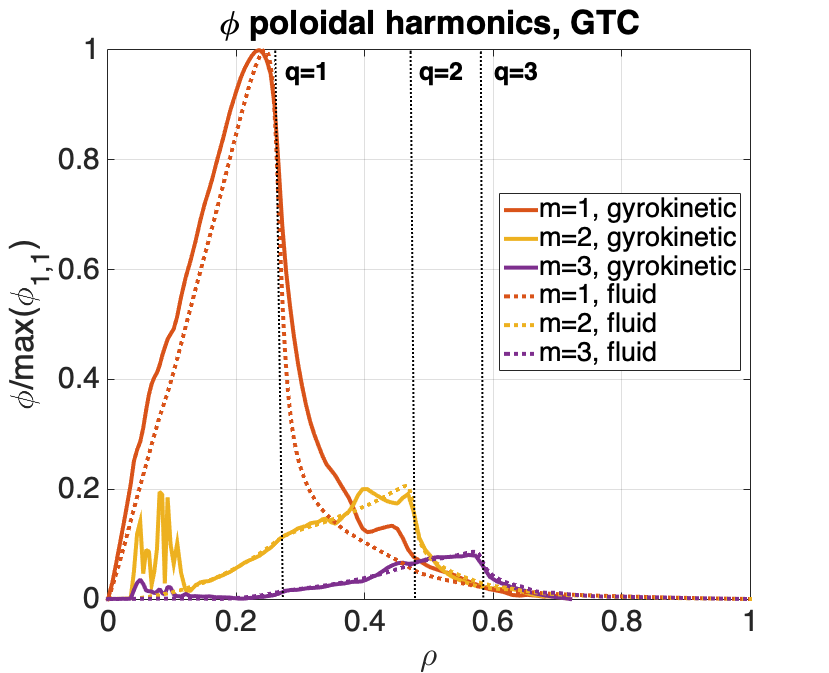}
   \caption{}
\end{subfigure}     
\begin{subfigure}{.49\textwidth} 
   \centering
   \includegraphics[scale=0.27]{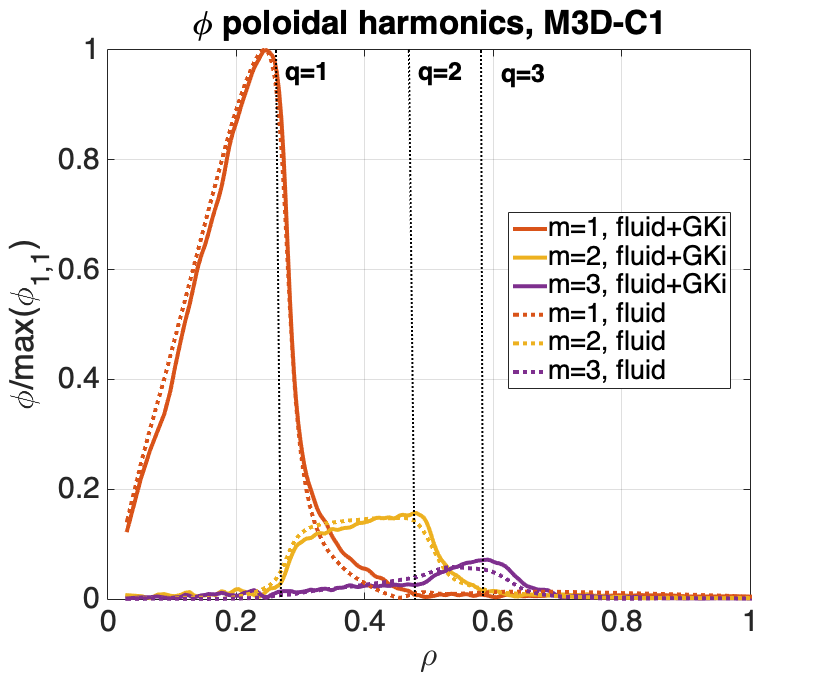}
   \caption{}
\end{subfigure}  
\caption{(a-b) Electrostatic potential of the internal kink in the poloidal plane and (c-d) poloidal harmonics of the n=1 electrostatic potential with gyrokinetic ions.}
\label{phi_GKi}
\end{figure}\\\\
The mode structure of the internal kink instability obtained from GTC and M3D-C1 simulations with kinetic thermal ions is shown in Figure (\ref{phi_GKi}). The simulation domain used for gyrokinetic simulations with GTC is slightly smaller than for fluid simulations, with $\psi_{edge}=0.8\psi_{lcfs}$ where $\psi_{edge}$ is the edge flux surface of the simulation domain. The kink growth rate decreases slightly due to the smaller edge flux surface, which is chosen  to avoid numerical instabilities at the edge plasma. The internal kink mode structure in both GTC and M3D-C1 simulations with kinetic thermal ions is similar to the one obtained in the ideal MHD limit. The overall mode structure in the poloidal plane is the same as in Figure (\ref{mode2D_bench}) (b-c). The m=1 harmonic of the electrostatic potential is however slightly broadened with kinetic thermal ions in GTC simulations according to Figure (\ref{phi_GKi}) (c).\\\\
The thermal ion temperature scan in Figure (\ref{gamma_GKi}) (a) was performed with and without the inclusion of trapped electron kinetic effects in GTC simulations. Trapped electrons were found to have no effect on the internal kink growth rate and frequency, nor on its mode structure. It can therefore be concluded that trapped electrons do not play a role in the internal kink linear dynamics.
\section{Conclusion and perspectives}
The DIII-D discharge \#141216 is used in this paper to carry out a verification and validation study on the internal kink instability for gyrokinetic and kinetic-MHD codes. The EFIT reconstruction of the experimental discharge is reprocessed through the equilibrium solver CHEASE in order to provide an identical input for all codes, which is crucial for a precise benchmark. An excellent quantitative agreement is obtained between all codes for the internal growth rate in the ideal MHD limit. The mode structures between all codes also agree very well, with a dominant n=m=1 harmonic emerging in all cases. Several physical elements are identified as essential to obtain a quantitative agreement between gyrokinetic and kinetic-MHD codes. The equilibrium current used in gyrokinetic codes needs to be calculated precisely by taking the curl of the magnetic field, the poloidal variations of the equilbrium current being often discarded in gyrokinetic codes, even though they have an important effect on low-n MHD modes. The second essential physical element is the inclusion of compressible magnetic fluctuations $\delta B_{\parallel}$. The internal kink is found stable in both GTC and M3D-C1 simulations when $\delta B_{\parallel}$ is artificially suppressed. This result is consistent with recent analytical work \cite{Graves2019} showing that neglecting the $\delta B_{\parallel}$ contribution has a strong stabilizing effect on the internal kink instability, to the extent that kink modes are unlikely to be recovered in the reduced MHD limit.\\Simulation results obtained from GTC and XTOR-K in the limit of ideal MHD are then validated against experimental measurements. The simulated $\delta T_e/T_e$ mode structures are found to agree quantitatively with the ECE measurements after adjustment of the q=1 position in the equilibrium reconstruction. The equilibrium reconstruction being subject to uncertainties, especially at the core plasma where experimental measurements are the most challenging, gyrokinetic/kinetic-MHD simulations can therefore be used to constrain some key parameters such as the q=1 position. \\ After the verification and validation of internal kink simulations in the ideal MHD limit, the kinetic effects of thermal ions and electrons are investigated using the initial value codes GTC, M3D-C1 and XTOR-K. Kinetic thermal ions are found to have stabilizing effects in kinetic-MHD simulations, where growth rates obtained between M3D-C1 and XTOR-K quantitatively agree with each other. Kinetic thermal ions are however destabilizing in gyrokinetic GTC simulations, due to an additional drive brought by the thermal ion temperature gradient and the finite parallel electric field. Trapped electrons are not observed to have any effect on the internal kink mode. \\ Future work focused on the fishbone phase of the DIII-D discharge will be conducted with realistic fast ion distributions. Both the present kink V\&V and the future fishbone V\&V constitute essential steps towards comprehensive GTC simulations describing realistically all channels for plasma confinement properties and energetic particle transport in burning plasmas. Such simulations will be able to accurately predict alpha particle transport, critical for  ITER experiments.
\section*{Acknowledgments}
The authors would like to thanks Liu Chen for helpful discussions. This work is supported by the U.S. Department of Energy (DOE) SciDAC project ISEP,  and used resources of the Oak Ridge Leadership Computing Facility at Oak Ridge Natinal Laboratory (DOE Contract No. DE-AC05-00OR22725) and the National Energy Research Scientific Computing Center (DOE Contract No. DE-AC02-05CH11231). It is also funded by the U.S. Department of Energy (DOE) under contract No. DE-AC02-09CH11466. This research used the Traverse cluster at Princeton University. This study is partially based upon work using the DIII-D National Fusion Facility, a DOE Office of Science user facility, under Awards DE-FC02-04ER54698 and DE-SC0020337.\\\\
This work has been carried out within the framework of the EUROfusion Consortium and has received funding from the Euratom research and training programme 2014-2018 and 2019-2020 under grant agreement No 633053. The views and opinions expressed herein do not necessarily reflect those of the European Commission. We benefited from HPC resources of  CINES from GENCI (projects no. 0510813) and the PHYMATH meso-center at Ecole Polytechnique.\\\\
This report was prepared as an account of work sponsored by an agency of the United States Government. Neither the United States Government nor any agency thereof, nor any of their employees, makes any warranty, express or implied, or assumes any legal liability or responsibility for the accuracy, completeness, or usefulness of any information, apparatus, product, or process disclosed, or represents that its use would not infringe privately owned rights. Reference herein to any specific commercial product, process, or service by trade name, trademark, manufacturer, or otherwise does not necessarily constitute or imply its endorsement, recommendation, or favoring by the United States Government or any agency thereof. The views and opinions of authors expressed herein do not necessarily state or reflect those of the United States Government or any agency thereof
\appendix
\section{Simulations models and setups}
\subsection{Gyrokinetic model}
\subsubsection{GTC}
GTC \cite{Lin1998}\cite{Holod2009}\cite{Xiao2015} is a global nonlinear gyrokinetic code employing a PIC module to describe gyrokinetic ions and electrons with both $\delta$f and full-f methods. Ion species are described with gyrokinetic equations, while the electron dynamics is obtained from the drift kinetic equation. The DKE can either be solved with a conservative scheme preserving the tearing parity \cite{Bao2017}, or with a fluid-kinetic formulation \cite{Holod2009} removing the tearing parity by expanding the perturbed electron distribution function into the adiabatic contribution at the lowest order, and non-adiabatic effects at the higher orders. The MHD limit can be recovered by neglecting ion kinetic effects and only considering adiabatic electrons. The perturbed electromagnetic field is obtained from the gyrokinetic Poisson equation and both parallel and perpendicular Amp\`ere's law, allowing to retain compressional magnetic perturbations $\delta B_{\parallel}$ \cite{Dong2017}. GTC has been originally used to study micro-turbulence in tokamak plasmas \cite{Lin1998} before being applied to Alfv\'en type meso-scales instabilities \cite{Deng2012a}\cite{Spong2012}\cite{Taimourzadeh2019} and more recently large scale kinetic-MHD instabilities \cite{McClenaghan2014}\cite{Liu2014}\cite{Liu2016}\cite{Shi2019}. In this paper, the fluid-electron model is used to solve the electron drift kinetic equation (DKE). At lowest order, with only adiabatic electrons, the limit of ideal full MHD is obtained in section 3. Higher orders are then retained in section 5 to recover kinetic effects of trapped electrons. Kinetic thermal ions are also included in section 5. All kinetic species are described using isotropic Maxwellian distributions without equilibrium rotation. The whole simulation domain is used for ideal MHD simulations, while simulations with kinetic species use $\rho\in[0,0.8]$ to avoid numerical instabilities. The time step size is t =0.045 $\tau_A$ , where  $\tau_A=V_A/R_0$ is the Alfv\'en time, with $V_A$ the Alfv\'en speed and $R_0$ the major radius. GTC uses Boozer coordinates and global field-aligned mesh \cite{Lin2002}\cite{Xiao2015} for the kink simulations. The radial and poloidal grid size spacings are respectively $\Delta r/\rho_i\sim 3.8$, $r\Delta\theta\rho_i\sim7.5$, where $\rho_i$ is thermal ion Larmor radius, and 24 grid points are used in the parallel direction. For each kinetic species, 200 marker particles are used per cell.
\subsection{Kinetic-MHD hybrid models}
\subsubsection{GAM-solver}
General Alfv\'enic Mode solver (GAM-solver) is a newly developed eigenvalue code for solving the dispersion relation of various drift-Alfv\'enic modes in experimental geometry. The perturbed fields in GAM-solver are represented using Fourier series expansion in poloidal and toroidal directions, and the finite difference method in the radial direction and spectral method in poloidal and toroidal directions are applied to construct the operator matrices. In current progress, the MHD module in GAM-solver builds on the reduced ideal MHD model using slow sound approximation \cite{Bao2021} and the drift and resistive MHD model with full plasma compressibility \cite{Bao2021}, which have been verified with Alfv\'en eigenmode, kink mode and kinetic ballooning mode physics \cite{Bao2021}. The gyrokinetic module in GAM-solver is based on well-circulating and deeply trapped approximations, which has been successfully applied to explain the energetic electron excitation of Alfv\'en eigenmodes observed in EAST experiments recently \cite{Zhao2021}.
For the simulation results of ideal kink mode in present work, the fully ideal MHD model in GAM-solver is used to be consistent with other codes on ideal kink mode physics, which includes the MHD vorticity equation with geodesic compressibility \cite{Bao2020}, and the parallel sound wave compressibility and  effect \cite{Bao2021}, while the diamagnetic and resistive effects are removed. The radial grid number is 500, and the poloidal Fourier harmonic number is 21 in the range from m=-10 to m = 10.
\subsubsection{M3D-C1}
M3D-C1 is a kinetic-MHD simulation code which can be used to study the interaction between MHD activities and kinetic effects brought by energetic particles. In this model, the bulk plasma is described using the MHD equations, and EPs are modeled following kinetic equations and described by markers. The MHD calculation is based on the finite-element MHD code M3D-C1 \cite{Jardin2012}\cite{Ferraro2009}\cite{Liu2021}, which solves extended-MHD equations using high-order finite-elements on an unstructured mesh with Cartesian coordinates. The particles are advanced following drift-kinetic equations using a slow manifold Boris algorithm \cite{Xiao2021}, which has been tested to show good long-time conservation property. The coupling with the MHD equation is conducted using either pressure coupling \cite{Fu2006} or current coupling  \cite{Todo1998}, where the pressure or current is calculated by integrating the moments of particles with the $\delta f$ method. A gyrokinetic model including  the finite Larmor radius (FLR) effect is also developed.\\\\ In the benchmark results presented in this paper, a 2D mesh with 16000 elements is used for the MHD calculation and 4106 particles are used for the kinetic simulation. The mesh has a high density near q=1 flux surface in order to resolve the mode structure of ideal kink instability. Kinetic particles are initialized with isotropic Maxwellian distribution functions.
\subsubsection{NOVA-K}
NOVA-K \cite{Gorelenkov1999}\cite{Gorelenkov2000} is a kinetic extension of the ideal MHD code NOVA \cite{Cheng1986}\cite{Cheng1992}. The NOVA code has been successfully applied earlier to m/n=1/1 ideal kink destabilization by bulk plasma pressure \cite{Cheng1992} where the code was benchmarked against several other ideal MHD codes. The perturbative approach to the ideal mode stabilization by energetic ions with the plasma rotation was considered in applications to JET NBI sawtooth stabilization \cite{Gorelenkov2000}\cite{Graves2003}. In NOVA-K approach to this problem the perturbative distribution function response to flat top kink mode plasma displacement allows to include the fast ion contribution through real part of their potential energy. Stabilizing effects on 1/1 internal kink mode by NBI hot ions was demonstrated for JET DT experiments with NBI heating in high performance discharges for beam ions and fusion alphas \cite{Gorelenkov2000}. It was found that the plasma sheared rotation in JET reduced the stabilizing effect of NBI ions on the growth rate of 1/1 internal mode. \\ In the benchmark exercise of this paper 151 and 128 number of points were used in radial and poloidal directions respectively. The number of poloidal harmonics was 24. 
\subsubsection{XTOR-K}
XTOR-K \cite{Lutjens2008}\cite{Lutjens2010}\cite{Brochard2020a} is a nonlinear Kinetic-MHD code originally developed for the study of macroscopic fluid instabilities in tokamak plasmas. It was notably applied for the simulation of  the internal kink mode \cite{Bondeson1992}\cite{Lutjens1992}, tearing modes \cite{Luetjens2001}, sawtooth cycling \cite{Halpern2011}\cite{Nicolas2012} and NTMs \cite{Maget2016}. In the code, the bulk plasma is described by a set of nonlinear resistive two-fluid MHD equations, solved implicitly with a Newton-Krylov algorithm using flux surface coordinates $(\psi,\theta,\varphi)$. \\ XTOR-K was later extended to include kinetic effects of multiple ion species with a kinetic PIC module. Kinetic macro-particles are described in 6D to retain wave-particle resonances with ions' gyrofrequency and to incorporate exactly ion FLR effects. A full-f method is used in XTOR-K to capture the kinetic-MHD dynamics over long non-linear hybrid simulations. Kinetic ions are pushed on the electromagnetic field computed from XTOR-K fluid module, with a Boris-Buneman scheme. The particle moments are projected on a orthogonal grid $(R,\varphi,Z)$. The particle advance is performed self-consistently with the MHD advance by injecting pressure and current kinetic moments $\textbf{P}_k, \textbf{J}_k$ into the MHD perpendicular momentum equation. The kinetic-MHD version of XTOR-K was recently used to study the alpha fishbone instability in ITER plasmas \cite{Brochard2020a}\cite{Brochard2020b}. \\ In the benchmark results presented in this paper, a (201,64,12) flux grid is used for the MHD module and a (401,12,401) cartesian grid for the kinetic module. A few hundred millions of macro-particles are employed to describe thermal ions. Kinetic particles are initialized with isotropic Maxwellian distribution functions.
\bibliography{kink_paper}{}
\end{document}